\documentclass{article}

\PassOptionsToPackage{numbers,sort&compress}{natbib}
\usepackage[preprint]{neurips_2024}

\usepackage[utf8]{inputenc} 
\usepackage[T1]{fontenc}    
\usepackage{hyperref}       
\usepackage{url}            
\usepackage{booktabs}       
\usepackage{amsfonts}       
\usepackage{subfig}         
\usepackage{nicefrac}       
\usepackage{microtype}      
\usepackage{xcolor}         

\usepackage{enumitem}       
\usepackage{amsmath}
\usepackage{amstext}        
\usepackage{doi}
\usepackage{graphicx}       
\usepackage{float}

\usepackage{multirow,setspace,caption}
\usepackage{tikz}
\usepackage{array}

\usepackage{booktabs, array, siunitx}
\newcolumntype{L}[1]{>{\raggedright\arraybackslash}p{#1}}
\newcolumntype{R}[1]{>{\raggedleft\arraybackslash}p{#1}}

\title{WAKESET: A Large-Scale, High-Reynolds Number Flow Dataset for Machine Learning of Turbulent Wake Dynamics}

\author{%
  Zachary Cooper-Baldock\\
  Flinders University\\
  \And 
  Paulo E. Santos \\
  Flinders University \\
  PrioriAnalytica \\
  \And
  Russell S.A. Brinkworth \\
  Flinders University \\
  \And
  Karl Sammut \\
  Flinders University \\
}

\begin{document}

\maketitle

\begin{abstract}
Machine learning (ML) offers transformative potential for computational fluid dynamics (CFD), promising to accelerate simulations, improve turbulence modelling, and enable real-time flow prediction and control—capabilities that could fundamentally change how engineers approach fluid dynamics problems. However, the exploration of ML in fluid dynamics is critically hampered by the scarcity of large, diverse, and high-fidelity datasets suitable for training robust models. This limitation is particularly acute for highly turbulent flows, which dominate practical engineering applications yet remain computationally prohibitive to simulate at scale. High-Reynolds number turbulent datasets are essential for ML models to learn the complex, multi-scale physics characteristic of real-world flows, enabling generalisation beyond the simplified, low-Reynolds number regimes often represented in existing datasets. This paper introduces WAKESET, a novel, large-scale computational fluid dynamics (CFD) dataset of highly turbulent flows, designed to address this critical gap. The dataset captures the complex hydrodynamic interactions during the underwater recovery of an autonomous underwater vehicle (AUV) by a larger extra-large uncrewed underwater vehicle (XLUUV). It comprises 1,091 high-fidelity Reynolds-Averaged Navier-Stokes (RANS) simulations, augmented to 4,364 instances, covering a wide operational envelope of speeds (up to Reynolds numbers of $1.09 \times 10^8$) and turning angles. This work details the motivation for this new dataset by reviewing existing resources, outlines the comprehensive hydrodynamic modelling and validation underpinning its creation, and describes its final structure. The dataset's focus on a practical engineering problem, its scale, and its high turbulence characteristics make it a valuable resource for developing and benchmarking ML models for flow field prediction, surrogate modelling, and autonomous navigation in complex underwater environments.
\end{abstract}


\section{Introduction}
\label{sec:Introduction}

Computational fluid dynamics (CFD) faces a fundamental challenge: high-fidelity simulations of practical engineering flows are extraordinarily expensive, often requiring weeks to months of computation on supercomputing clusters for a single flow condition. This computational burden severely limits design space exploration, real-time control applications, and uncertainty quantification—all critical needs in modern engineering \cite{BruntonEtAl_2019, VinuesaAndBrunton_2022, BruntonSpecialIssue}. The problem is particularly acute for turbulent flows at high Reynolds numbers, which characterise most real-world applications from aircraft aerodynamics to underwater vehicle manoeuvring. Direct Numerical Simulation (DNS), which resolves all turbulence scales, remains computationally intractable for these problems, while lower-fidelity approaches like Reynolds-Averaged Navier-Stokes (RANS) sacrifice accuracy and often fail to capture complex flow phenomena reliably.

Machine learning (ML) offers a promising path forward to address these challenges. ML models can potentially serve as fast surrogate models that provide near-instantaneous flow predictions, enable data-driven turbulence closure models that improve accuracy, facilitate real-time flow control and optimisation, and reduce the dimensionality of complex flow fields for efficient analysis. However, realising this potential requires a fundamental prerequisite that the CFD community currently lacks: large, diverse, high-fidelity training datasets that capture the complexity of real engineering flows.

This need is particularly critical for highly turbulent, high-Reynolds number flows. Unlike many other fields where ML has been rapidly adopted—such as computer vision, which benefits from massive datasets like ImageNet \cite{ImageNet} containing millions of labelled examples—fluid mechanics suffers from a severe data scarcity problem. The computational expense of generating high-quality CFD data, combined with the high dimensionality and complexity of flow fields, has resulted in existing datasets that are often limited in scope, size, and diversity \cite{BruntonEtAl_2019, VinuesaAndBrunton_2022}. Most available datasets focus on canonical flows, low Reynolds numbers, or two-dimensional simplifications, leaving a critical gap for training ML models capable of handling practical engineering problems involving complex three-dimensional turbulent flows.

This paper directly addresses this gap by introducing WAKESET: a novel, large-scale CFD dataset specifically designed for training and benchmarking ML models in highly turbulent flow regimes. WAKESET represents a significant advancement over existing resources, providing 4,364 high-fidelity three-dimensional flow field instances spanning Reynolds numbers up to $1.09 \times 10^8$—among the highest in publicly available CFD datasets (Table \ref{tab:CFD-Datasets}). The dataset is grounded in a practical engineering application: the underwater recovery of an autonomous underwater vehicle (AUV) by an extra-large uncrewed underwater vehicle (XLUUV), a scenario that generates rich, complex hydrodynamic phenomena including boundary layer interactions, propeller wakes, payload bay recirculation zones, and maneuver-induced vortical structures.

The development of WAKESET follows a rigorous two-phase approach. First, a comprehensive hydrodynamic analysis of the XLUUV recovery scenario was conducted to identify critical flow phenomena, validate the CFD methodology, and establish the foundation for dataset generation—work that has been published in both conference proceedings \cite{SUBSTEC} and journal form \cite{ZacIEEE}. Second, this specific analysis was systematically generalised and expanded to create a broad, versatile dataset suitable for training robust ML models. This generalisation involved developing a representative XLUUV geometry, expanding the parameter space to cover diverse operational conditions (speeds from 0.10 to 5.00 m/s and turning angles from 0 to 60 degrees), and employing data augmentation techniques to further increase dataset size and diversity.

The primary contribution of this work is the provision of a comprehensive, publicly available dataset that addresses the critical needs identified by the fluid mechanics ML community: high Reynolds numbers to represent practical turbulent flows, three-dimensional data to capture real flow physics, sufficient instances (4,364) to train data-intensive deep learning architectures, varied flow conditions to promote model generalisation, and a practical engineering focus to ensure relevance. By providing this resource, we aim to accelerate progress in ML applications for fluid mechanics, enabling the development of models that can handle the complexity of real-world engineering flows and ultimately help overcome the computational bottlenecks that have long constrained CFD-based design and analysis.

The remainder of this paper is organised as follows: Section \ref{sec:Motivation} provides a detailed motivation by examining the limitations of existing CFD datasets and the specific requirements for effective ML training in fluid mechanics. Section \ref{sec:InitialHydro} introduces the XLUUV recovery scenario as a challenging and relevant case study. Section \ref{sec:Methodology} describes the comprehensive methodology for dataset creation, from the initial hydrodynamic analysis to the final generalised dataset. Section \ref{sec:WAKESET} presents the WAKESET dataset characteristics, augmentation strategy, and availability. Section \ref{sec:Benchmarks} provides performance benchmarks using state-of-the-art generative models. Finally, Sections \ref{sec:Discussion} and \ref{sec:Conclusion} discuss the implications and conclusions of this work.

\section{Motivation}
\label{sec:Motivation}

The computational demands of high-fidelity CFD simulations represent one of the most significant bottlenecks in modern engineering design and analysis \cite{BruntonEtAl_2019, BruntonSpecialIssue, VinuesaAndBrunton_2022}. For practical engineering problems involving complex geometries and turbulent flows at high Reynolds numbers, a single RANS simulation can require hours to days of computation on high-performance computing clusters, consuming thousands of CPU-hours. More accurate approaches, such as Large Eddy Simulation (LES), can increase these requirements by orders of magnitude, while Direct Numerical Simulation (DNS)—the gold standard for resolving all turbulence scales—remains computationally prohibitive for all but the simplest flows at moderate Reynolds numbers. A single DNS of a practical engineering geometry at realistic operating conditions could require millions of CPU-hours and generate petabytes of data \cite{SCA24}.

This enormous computational cost has profound implications. Design optimisation, which requires evaluating hundreds or thousands of design variants, becomes impractical when each evaluation takes days. Uncertainty quantification, which demands Monte Carlo sampling across parameter spaces, is similarly constrained. Real-time flow prediction for control applications is impossible when simulations cannot keep pace with physical time. Parametric studies to understand flow physics across operating conditions remain limited to sparse sampling of the parameter space. These limitations fundamentally constrain what engineers can achieve with CFD, forcing reliance on simplified models, reduced-order approaches, or expensive experimental testing.

Beyond computational cost, the technical complexity of CFD presents additional challenges. High-fidelity simulations require specialized expertise in mesh generation, turbulence modeling, numerical methods, and convergence analysis. Setting up a single simulation for a new geometry can take weeks of expert time. Ensuring mesh independence and solution convergence demands systematic studies that multiply the computational burden. The need for domain expertise limits who can effectively use CFD tools and slows the iteration cycles essential for innovation. These factors combine to create a substantial barrier between engineering problems and their fluid dynamic solutions.

The integration of ML techniques into CFD offers significant potential for overcoming these longstanding challenges. These challenges, as highlighted by seminal reviews \cite{BruntonEtAl_2019, BruntonSpecialIssue, VinuesaAndBrunton_2022}, include not only the computational cost of high-fidelity simulations, but also difficulties in turbulence modelling and closure, the curse of dimensionality in flow data analysis, and the complex control of unsteady flows. ML approaches promise advancements in areas such as the development of accurate surrogate models that can predict flow fields orders of magnitude faster than traditional CFD, discovering turbulence closure models from data that improve RANS accuracy, identifying coherent structures and reduced-order representations of complex flows, and enabling real-time flow control strategies. However, a major obstacle hindering progress is the scarcity of large, high-fidelity datasets suitable for training robust and generalisable ML models.

This contrasts sharply against other fields such as computer vision, where readily available and extensive datasets (e.g., MNIST \cite{MNIST}, EMNIST \cite{10_GMSEGeneralPaper}, CIFAR \cite{CIFAR10}, ImageNet \cite{ImageNet}) have fuelled rapid advancements both architecturally and on the application front. These image datasets each contain thousands to millions of labelled examples, enabling the training of deep neural networks with vast numbers of parameters. This scale and diversity are crucial for mitigating the risk of over-fitting and allowing such complex models to learn generalisable features. Similarly, in fluid mechanics, larger and more diverse CFD datasets are essential for training sophisticated ML architectures that can capture complex flow physics without memorisation of the training data. 

The fluid mechanics community faces a fundamentally different landscape. Generating high-quality CFD data is resource intensive. High-fidelity simulations, particularly those involving complex geometries and turbulent, high Reynolds number, flows, demand substantial computational power and time \cite{SCA24}. Direct Numerical Simulation (DNS), considered the "gold-standard" for resolving all scales of turbulent motion, is computationally prohibitive, if not impossible, for most practical engineering problems, often requiring weeks or months of computation on super-computing clusters for a single flow condition. Experimental data acquisition, while valuable, is also constrained by the high costs and logistical complexities associated with the sophisticated experimental measurement techniques like particle image velocimetry (PIV) required to resolve the flow field. 

Due to these limitations, existing CFD datasets commonly used for ML in fluid mechanics are often limited in scope, size and diversity. The John Hopkins Turbulence Database (JHTDB) \cite{JHTDB}, while valuable for fundamental turbulence research, provides data primarily for canonical flows (e.g., isotropic turbulence, channel flow) and lacks the geometric and boundary condition variations required in many engineering applications. Datasets associated with specific research models, such as the NASA Common Research Model \cite{NASA_Data}, are often focused on particular configurations and not sufficiently extensive for training generalisable ML models. 

Table \ref{tab:CFD-Datasets} provides a comparative overview of publicly available CFD datasets, highlighting their limitations. Notably, many 3D flow datasets, which are essential for capturing the full complexity of real-world flows, are limited to a few hundred instances or less \cite{WindsorML, DrivAerML, AhmedML, BLASTNet, JHTDB, McConkey}. This limited scale often restricts the complexity of ML models that can be trained effectively, potentially leading to under-fitting of the complex fluid dynamics and overfitting to the small dataset, thereby hampering the development of performant and generalisable models. Furthermore, many datasets are generated at fixed flow speeds \cite{WindsorML, DrivAerML, MegaFlow2D, AhmedML}, limiting their ability to train models that can generalise across a range of Reynolds numbers (speeds). Many datasets are also restricted to 2D flows \cite{MegaFlow2D, AirfRANS, Huang}, which, while useful for some applications, do not fully represent the three-dimensional nature of most engineering problems. Finally, several datasets are focused on relatively low-speed, less turbulent flows, often emphasising DNS-level accuracy over the practical, high-Reynolds number regimes relevant to many engineering applications \cite{MegaFlow2D, BLASTNet, JHTDB}. This, in turn, results in datasets with a limited number of instances, potentially insufficient for training data-intensive ML architectures. 

\begin{table}[H]
\caption{Overview of publicly available CFD datasets commonly used for ML training.  The table highlights the year of publication, the number of instances, the Reynolds number range, whether the velocity is fixed, the dimensionality of the data, and the fluid model used for simulation. *34 simulations with 700 time steps each. **10 simulations with 15,791 time steps each.}
\label{tab:CFD-Datasets}
\begin{center}
\footnotesize
\begin{tabular}{lllllll}
\toprule \toprule
\textbf{Dataset}             & \textbf{Year} & \textbf{Instances} & \textbf{Reynolds}    & \textbf{Fixed Velocity} & \textbf{Dimension} & \textbf{Fluid Model} \\ \toprule \toprule \addlinespace
WindsorML \cite{WindsorML}   
& 2025 
& 355       
& $2.9 \times 10^{6}$            
& Yes            
& 3D        
& WMLES       
\\ 
\addlinespace
\hline
\addlinespace
DrivAerML \cite{DrivAerML}   
& 2024 
& 500       
& $7.2 \times 10^{6}$            
& Yes            
& 3D        
& RANS-LES    
\\ 
\addlinespace
\hline
\addlinespace
MegaFlow2D \cite{MegaFlow2D} 
& 2023 
& 3000      
& 300                            
& Yes            
& 2D        
&            
\\ 
\addlinespace
\hline
\addlinespace
AhmedML \cite{AhmedML}       
& 2024 
& 500       
& $7.7 \times 10^{5}$            
& Yes            
& 3D        
& LES         
\\ 
\addlinespace
\hline
\addlinespace
BLASTNet \cite{BLASTNet}     
& 2023 
& 34*       
& 8000                           
& No             
& 3D        
& DNS         
\\ 
\addlinespace
\hline
\addlinespace
JHTDB \cite{JHTDB}           
& 2008 
& 10**      
& 1000                           
& No             
& 3D        
& DNS         
\\ 
\addlinespace
\hline
\addlinespace
McConkey et al. \cite{McConkey} 
& 2021 
& 29     
& 27,850                         
& No             
& 2D, 3D    
& RANS        
\\ 
\addlinespace
\hline
\addlinespace
AirfRANS \cite{AirfRANS}     
& 2022 
& 1,000     
& $6 \times 10^{6}$             
& No            
& 2D        
& RANS        
\\ 
\addlinespace
\hline
\addlinespace
Huang \cite{Huang}           
& 2020 
& 30,000    
& -                              
& No             
& 2D        
& LES         
\\ 
\addlinespace
\hline
\addlinespace

\textbf{WAKESET}                
& \textbf{2025} 
& \textbf{4,364}     
& $\mathbf{1.09 \times 10^{8}}$           
& \textbf{Yes}             
& \textbf{3D}        
& \textbf{RANS}        
\\ \addlinespace \toprule \toprule
\end{tabular}
\end{center}
\end{table}

The limitations in existing CFD datasets stem from three primary factors:

\begin{enumerate}
    \item \textbf{Computational Cost:} High-fidelity simulations, especially for turbulent flows, are computationally expensive. Generating thousands of simulations is often impractical due to economic, computational and logistical constraints.

    \item \textbf{Dimensionality and Complexity:} CFD data is inherently high-dimensional, typically comprising three-dimensional fields of velocity, pressure and other relevant variables, often evolving over time. Storing, managing and processing such data requires significant infrastructure and computational resources.

    \item \textbf{Data Diversity:} Achieving sufficient diversity across a broad range of flow conditions, geometries and boundary conditions is essential for training ML models that can generalise effectively to unseen scenarios. This requires a systematic exploration of the parameter space, which can be computationally demanding. 
\end{enumerate}

To enable effective training of ML models for fluid mechanics, there is a clear need for new datasets that address these limitations. Specifically datasets should encompass several key aspects. The first is the inclusion of high Reynolds number, sufficiently turbulent flows. This is required to represent the conditions encountered in practical engineering applications which have high degrees of turbulence. The second is the capture of 3D, or ideally 4D, data. This is required to ensure that ML architectures are able to train at the same dimension as real flows. Thirdly, a sufficient number of instances needs to be provided, to enable exploration of data-driven modelling techniques - such as those seen in computer vision. Finally, varied flow conditions, parameters, or geometries should be used to assess the generalisation of different model types. This is required to ensure that over-fitting is not occurring. Any developed dataset should also seek to incorporate a sufficient degree of accessibility and standardisation, making the data readily available to the research community in a standardised format will facilitate benchmarking and collaboration. The development of such datasets will pave the way for advancements in ML for fluid mechanics, hopefully mirroring the progress seen in computer vision and natural language processing (NLP), aided by their dataset availability.

The primary objective of this work is to create a comprehensive CFD dataset that addresses these critical needs. The research focuses on the complex and practically relevant scenario of underwater vehicle recovery, specifically the berthing of an AUV within an XLUUV. This scenario presents a rich set of hydrodynamic challenges and effects, including complex flow interactions, turbulence, and varying boundary conditions, making it ideal for generating a diverse and challenging dataset.

\section{XLUUV Recovery: A Challenging and Relevant Case Study}
\label{sec:InitialHydro}

The underwater recovery of AUVs presents significant hydrodynamic challenges, particularly when it involves berthing within an XLUUV. This scenario is characterised by complex flow interactions arising from the presence of the large mothership vessel, its propeller induced wakes, and any manoeuvring actions. A thorough understanding and accurate modelling of these flow-field phenomena are essential for ensuring the safety and efficiency of AUV deployment and retrieval, especially in situations where the use of surface vessels is impractical or undesirable.

While various methods for underwater vehicle recovery exist \cite{ZacIEEE, SUBSTEC}, including tethered systems, seafloor capture systems, and surface vessel-based mechanisms, these approaches are often unsuitable for challenging, hazardous, or covert operations. The advent of larger autonomous underwater vehicles offer a promising solution through the entirely underwater launch and recovery of smaller AUVs via a payload bay located onboard an XLUUV mothership. While the launch of AUVs from submarines is well understood, subsequent underwater recovery, particularly using torpedo tubes, poses considerable operational, hydrodynamic, and control challenges. Recent XLUUV designs, such as the Boeing Echo Voyager, feature payload bays sufficiently large to accommodate a variety of smaller AUV platforms, including those like the GRAALtech X300, Remus 300/600, and the Kingfish vehicles. The capability to launch and recover smaller AUVs from an XLUUV would directly enable extended searches, reduce operational cost and risk, and enhance mission flexibility. 

However, ensuring safe and reliable berthing operations within the XLUUV's payload bay requires a detailed understanding of the hydrodynamic phenomena involved, particularly in the vicinity of the payload bay opening, the boundary layer and the wake region of the XLUUV. While some studies have investigated AUV docking with conventional static, externally attached, or towed docks, these have largely focused on conventional submarine or surface vessel geometries. No existing prior work, at the time of writing, had specifically addressed the unique hydrodynamic challenges of AUV berthing within an XLUUV’s internal payload bay.

The XLUUV recovery scenario was selected as the case study for this research due to its inherent complexity, practical relevance, and the critical need for a deeper understanding of the associated hydrodynamic phenomena. This scenario involves intricate flow interactions that are representative of the real-world challenges encountered in underwater vehicle operations. The complex interaction between the host vessel, propeller wakes, and manoeuvring actions provides a deep and diverse set of flow conditions, making it ideal for generating a dataset for training robust and generalisable ML models. The focus on underwater vehicles directly provides a practical engineering context, whilst also seeking to develop a valuable dataset that will advance the application of ML more broadly within fluid mechanics. The insights derived from this analysis will be used to inform the subsequent generalisation and dataset development steps, ensuring that the final dataset captures the essential hydrodynamic features necessary for training effective ML models.

\section{Methods}
\label{sec:Methodology}

The creation of the WAKESET dataset was a two-phase process. First, a detailed hydrodynamic analysis of a specific subset of flow conditions was undertaken, as detailed comprehensively in our existing work \cite{ZacIEEE}. This foundational study served to identify critical flow phenomena, validate the CFD methodology and numerical results, and provide crucial insights into the hydrodynamic interactions. Second, building on these findings, the simulation was systematically generalised and scaled to generate a broad, versatile and large-scale dataset suitable for training and evaluating a wide range of machine learning models. This section details both phases of this process, from the initial model setup and analysis to the final dataset's structure and characteristics.

\subsection{Foundational Hydrodynamic Analysis}
\label{sec:Foundational}

To create the comprehensive dataset, a detailed hydrodynamic analysis of the XLUUV recovery scenario was undertaken. This section outlines the methodology used to develop and validate the initial CFD model, including the selection of vehicle geometries, approach parameters, hydrodynamic phenomenon of interest and the setup and discretisation of the simulation for computation and subsequent analysis. The choices made in each aspect of the modelling process were carefully considered to ensure that the resulting data would be both representative of the complex flow physics involved with XLUUV recovery and suitable for extension into a training dataset. 

\subsubsection{Vehicle Geometries}
\label{sec:VehicleGeometry}

The initial CFD model incorporated two vehicles: an XLUUV acting as the host vessel and a smaller AUV representing the vehicle being recovered. For the XLUUV, a representative model was developed based on common design features observed in existing XLUUV platforms. The initial XLUUV model has a length of 26 meters and includes a large centralised payload bay. For the AUV, a generic model was developed, drawing structural inspiration from the GRAALtech X300. The AUV model features a simplified hull structure with a length of 2.2 meters and a diameter of 0.15 meters. More specifics are available in \cite{ZacIEEE}. 

\subsubsection{Approach Paths, Trajectories and Speeds}
\label{sec:Paths}

A critical aspect of the initial hydrodynamic analysis was to explore a diverse set of AUV approach conditions. The AUV's approach to the XLUUV was systematically varied in terms of its path, trajectory and speed. Two variations of the approach process were considered: a flow-aligned approach and a path-aligned approach. For each of the two approach configurations, three distinct trajectories were analysed, involving varying the slope of the ascent. Three speeds were assessed for each configuration and trajectory: 1 knot (0.51 m/s), 2 knots (1.03 m/s), and 3 knots (1.55 m/s) the justification for which is provided in \cite{ZacIEEE}.

\subsubsection{Hydrodynamic Phenomenon of Interest}
\label{sec:HydrodynamicPhenomena}

Several key hydrodynamic phenomena were analysed to understand their impact on the AUV during the recovery process. The full details of this are provided in Appendix \ref{App:HydrodynamicPhenomenon}. The lift and drag coefficients, $C_L$ and $C_D$, respectively, were assessed to understand the forces acting on the AUV throughout the approach. It is particularly important to understand the changes in force with respect to the 3D position of the XLUUV wake structure.

To quantify the turbulence experienced during navigation into the payload bay, the turbulence intensity, turbulent kinetic energy and vorticity magnitude were assessed. The turbulence intensity ($I$) details the fluctuations of velocity with respect to the magnitude of the freestream velocity condition, giving an indication of the magnitude of turbulent flow oscillation. The turbulent kinetic energy ($k$) details the energy that is exerted by the turbulent regions of the flow field. The vorticity magnitude ($\vec{w}$) details the rotational vorticity of the flow field, as experienced by the AUV as it navigates along the desired trajectories. The equations and a more detailed explanation of these metrics is provided in Appendix \ref{App:HydrodynamicPhenomenon}.

\subsubsection{CFD Model Setup and Validation}
\label{sec:CFDModel}

The simulations were based on an operational depth of 100 meters. The water properties at this depth were based on the BRAN2020 \cite{BRAN202} ocean analysis for the Australasian region. The CFD simulations were performed using ANSYS Fluent 2021R2, employing the realizable $k-\epsilon$ turbulence model. The propellers of the XLUUV and AUV were modelled using the virtual blade model (VBM) based on an INSEAN E1619 propeller. A mesh convergence analysis was performed, and a grid size of 10 million cells was selected, as detailed in Appendix \ref{App:HydrodynamicModelGeneralisation}. The numerical results were validated against existing literature that was both computational and experimental. These are provided in more detail in \cite{ZacIEEE}.

\subsection{Key Insights from the Foundational Analysis}
\label{sec:InitialAnalysisFindings}

The results of the initial study \cite{ZacIEEE} provided critical insights that guided the design of the final, large-scale dataset. The analysis revealed the formation of a strong shear layer at the payload bay entrance and complex recirculation zones within it. This underscored the need for high mesh resolution in the payload bay and wake regions to capture these critical features in the final dataset.

The analysis of lift and drag coefficients identified regions of high hydrodynamic interaction, particularly near the XLUUV's propeller and at the payload bay entrance. The forces on the AUV were shown to be highly dependent on its trajectory and proximity to these features. This informed the need to accurately model the propeller's influence and ensure sufficient simulation density in these dynamic regions. The study demonstrated that turbulent characteristics (intensity, TKE, vorticity) are not uniform and are strongly influenced by the AUV's path and the XLUUV's wake. This complex, non-linear behaviour necessitated the creation of a dataset with a wide range of speeds and angles to capture these effects for robust ML model training.

\subsection{Generalisation to an ML Training Set}
\label{sec:Generalisation}

This section details the transition from the specific hydrodynamic analysis of XLUUV recovery (detailed in Section \ref{sec:Foundational}) to a generalised, comprehensive CFD dataset suitable for training machine learning models. The insights gained from the initial investigation - particularly regarding critical flow regions (such as the payload bay shear layer and the propeller wake), the sensitivity of the hydrodynamic forces to vehicle orientation and proximity, and the range of turbulent effects encountered - were instrumental in guiding the design and scope of this generalised dataset. Its development was thus guided by these findings and the overarching need to create a resource representative of diverse real-world hydrodynamic complexities, suitable for advancing ML applications in fluid mechanics. Key aspects of this process, directly informed by the preceding analysis, include the generalisation of the XLUUV model geometry, the expansion and parametrisation of flow conditions, the setup of the CFD model for large-scale dataset generation, and a description of the dataset's characteristics and availability.

\subsubsection{Generalisation of the XLUUV Model}
\label{sec:GeneralisationGeometry}

To ensure the dataset's broad applicability to various XLUUV designs and to avoid overfitting to the specific features of a single platform, a generalised XLUUV model was developed. This involved creating a reference model that captures the common design characteristics of multiple existing XLUUV platforms while maintaining a realistic representation of the hydrodynamic interactions relevant to a wide range of recovery scenarios and flow field conditions. The generalisation process focused on the following key aspects. The generalised model incorporates a rounded nose and flattened hull sides, top, and bottom, similar to most current XLUUV designs. This shape simplifies the incorporation of a payload bay and reduces flow separation compared to more rounded hulls. The complexities of some XLUUV's surface geometries, where communications structures are mounted, are simplified for enhanced generalisation. 

\begin{figure}[H]
\centering
\includegraphics[scale=0.95]{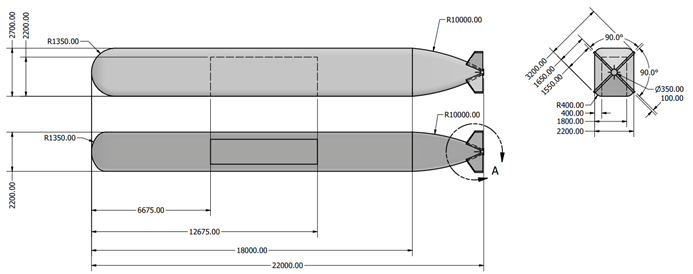}
\caption{XLUUV hull structure in generalised form. All dimensions are provided in millimetres (mm). Payload bay geometry shown in the centre of the XLUUV body. X fin tail design with rear propeller appended to the centre of this structure as an actuator disk model within ANSYS Fluent.}
\label{fig:XLUUV_Generalised}
\end{figure}

The model's overall dimensions were chosen to be representative of the larger XLUUV platforms currently in development, as these are more likely to be used for AUV recovery operations. A length of 22 meters was selected, falling within the range of existing designs (e.g., the Boeing Orca at 26 meters and the Klavesin at 6.5 meters \cite{35}). The beam was set at 2.2 meters x 2.7 meters, consistent with the typical cross-sectional diameters of 1.5 to 2 meters observed in various platforms \cite{11}. The payload bay design is crucial, as it significantly influences the flow patterns during AUV recovery. The generalised model incorporates a payload bay with an aspect ratio (length/width) of approximately 4, matching that of the Boeing Orca \cite{11}. The payload bay is located at the bottom centre of the hull, approximately 3 meters ahead of the stern, with its centre located 7 meters from the bow. The dimensions of the payload bay are 1.4 meters (width) x 2.2 meters (depth) x 6.0 meters (length). These dimensions and location were selected to ensure the dataset captures the essential flow features associated with internal payload bay recovery, while remaining generalisable. This geometry is provided in Figure \ref{fig:XLUUV_Generalised}.

The development of a generalised XLUUV model was crucial for creating a dataset with broad applicability. By focusing on common design features and avoiding platform-specific details, the model ensures that the dataset is relevant for training ML models that can generalise well across different XLUUV configurations, potentially. This generalisation step moves beyond the specific scenario analysed in \cite{ZacIEEE} and allows for the creation of a dataset with far broader utility. Furthermore, such a generalised dataset can serve as an excellent foundation for pre-training robust ML models. These pre-trained models could then be fine-tuned efficiently for specific vehicle designs using smaller, custom datasets via transfer learning \cite{AugmentedDataRotation}, significantly reducing the data generation burden for bespoke applications. 

\subsubsection{Parametrisation: Expanding the Flow Regime}
\label{sec:GeneralisationExpansion}

To capture a wide range of flow conditions and ensure the dataset's diversity, key parameters related to the XLUUV's motion were systematically varied, significantly expanding beyond the initial analysis. These parameters include the forward speed and the turning angle, which significantly influence the flow field around the XLUUV and, consequently, the hydrodynamic forces acting on any nearby object (such as an AUV during recovery). 

The forward speed of the XLUUV was varied from 0.10 m/s to 5.00 m/s in increments of 0.01 m/s. This range significantly exceeds typical docking speeds, ensuring the dataset encompasses a wide range of flow conditions, including general operation of the XLUUV, inline with standard operational speeds. The upper limit of 5.00 m/s was chosen to allow the model to be used in higher-speed operations and applications, should underwater berthing or manoeuvring technology mature to that point. This limit is above the typical operational speed of smaller AUV platforms (often limited to speeds less than 2.57 m/s \cite{40}) but provides a safety margin and ensures the dataset's applicability to a broader range of scenarios. Importantly, this range extends beyond the maximum speed (1.55 m/s) tested in the initial hydrodynamic analysis \cite{ZacIEEE}, creating a much more comprehensive dataset.

The turning angle of the XLUUV was varied from 0 degrees (dead ahead) to 60 degrees in increments of 5 degrees, representing a wide range of manoeuvring conditions, including turns beyond anticipated operational conditions. The inclusion of turning angles is crucial as it introduces lateral flow components that significantly affect the wake structure and the flow field around the payload bay, by excitation of strong vortical structures. This is especially important for AUV recovery, as the approaching vehicle may encounter complex, asymmetric flow patterns during turns.  The initial analysis \cite{ZacIEEE} focused on straight-line motion (0-degree turning angle); this parametrisation significantly expands the scope of the dataset.

To relate the forward speed, lateral speed, and turning angle, \ref{eqn:VelocityRelation} was used. This relates the total velocity of the XLUUV ($v_{tot}$) to the forward ($v_{x}$) and lateral velocity ($v_{y}$) components via the turning angle ($\theta$). 

\begin{equation}
\label{eqn:VelocityRelation}
v_{tot} = \sqrt{v_{x}^{2} + v_{y}^{2}} = \sqrt{(v\cos\theta)^{2} + (v\sin\theta)^{2}}
\end{equation}

Eq. (\ref{eqn:VelocityRelation}) provides a relationship between the total velocity, its components, and the turning angle. It was implemented in the CFD model to ensure that the simulated flow conditions accurately represented the XLUUV's motion for each combination of forward speed and turning angle. These relationships were parameterised within the CFD model to systematically generate the dataset, ensuring consistency and accuracy across all simulations.

The chosen ranges and increments for speed and turning angle, along with the governing equation, ensure that the dataset covers a wide operational envelope and captures the complex flow physics associated with XLUUV manoeuvres. This parametrisation is crucial for training ML models that can accurately predict hydrodynamic forces and flow fields under diverse operating conditions, including scenarios that may not be commonly encountered in typical AUV recovery operations, and significantly expands upon the initial hydrodynamic analysis.

\subsubsection{CFD Model for Large-Scale Generation}

The CFD model was adapted for large-scale data generation. This included a new mesh convergence study for the generalised geometry, resulting in the selection of a 14-million-cell mesh. An iteration convergence study was also performed, leading to a conservative cutoff of 300 iterations per simulation to ensure well-converged and reliable data.

\section{WAKESET}
\label{sec:WAKESET}

The WAKESET dataset consists of 1,091 individual CFD simulations, covering a wide range of flow conditions defined by the combinations of forward speeds (0.10 m/s to 5.00 m/s in increments of 0.10 m/s) and turning angles (0 degrees to 60 degrees in increments of 5 degrees).  The dataset generation was performed using ANSYS Fluent 2023R1 on the GADI supercomputer \footnote{Conducted during Q1 and Q2 2023, under an Australian Higher Performance Computing - Artificial Intelligence (HPC-AI) grant awarded by the National Computational Infrastructure.}. Each simulation was run for 300 iterations, ensuring convergence based on the iteration convergence study (Appendix \ref{App:HydrodynamicModelGeneralisation}).

\subsection{Dataset Characteristics}
\label{sec:DatasetCharacteristics}

Each simulation in the dataset provides a comprehensive set of flow field data, including:

\begin{itemize}
    \item Cell coordinate data (cell ID, location (x, y, z))
    \item Total pressure
    \item Absolute pressure
    \item Dynamic pressure
    \item Velocity magnitude ($v_{mag}$)
    \item Velocity components ($u, v, w$)
    \item Vorticity ($\omega$, computed via curl of velocity)
    \item Turbulence Intensity ($I$, computed via $k$ and $|\mathbf{v}|$)
\end{itemize}

Each simulation is decomposed into three separate files, each of which is denoted by the following filename convention:

\texttt{Forward\_[VVVV]\_ms\_Angle\_[AA]\_[TYPE]\_[SUFFIX]}

\begin{itemize}
    \item \textbf{VVVV}: Forward speed in mm/s (e.g., \texttt{5000} denotes $5.00$ m/s).
    \item \textbf{AA}: Yaw angle in degrees (e.g., \texttt{20}).
    \item \textbf{TYPE}: Geometry type (\texttt{CUBE\_128}, \texttt{VERTPLN}, \texttt{HORZPLN}).
\end{itemize}

The \texttt{CUBE\_128} files contain 3D flow fields interpolated onto a regular $128^3$ Cartesian grid, specifically optimised for volumetric machine learning tasks (e.g., 3D-GANs). The \texttt{PLN} files contain unstructured mesh data on planar slices, retaining the original CFD mesh density to facilitate high-fidelity boundary layer analysis.

\begin{figure}[H]
\centering
\includegraphics[scale=0.42]{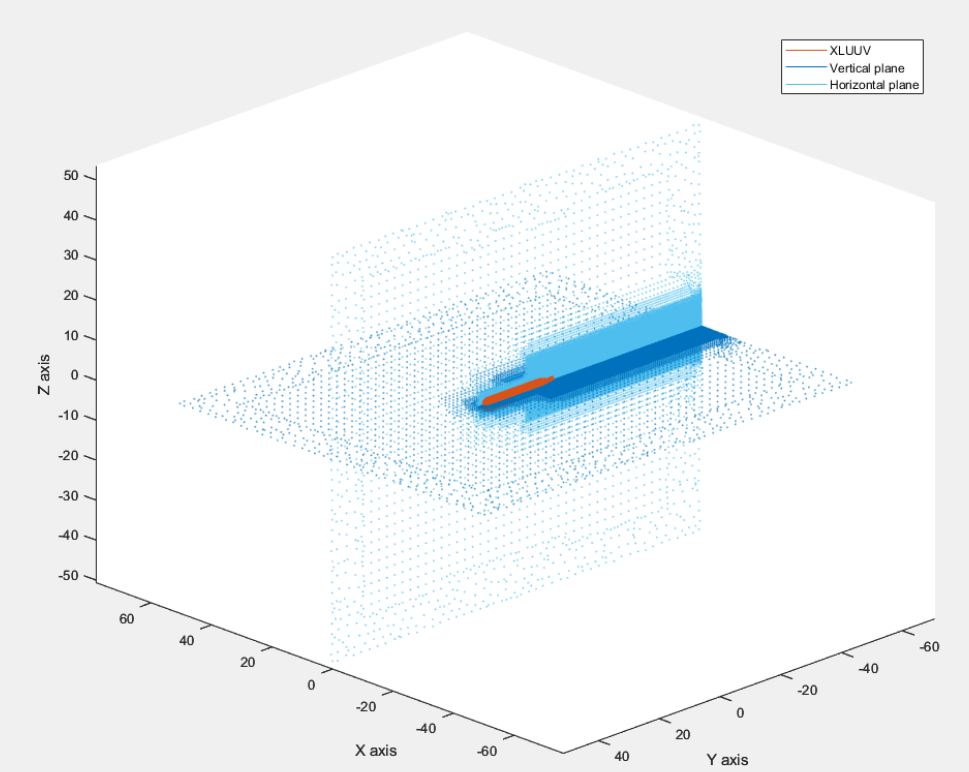}
\caption{XLUUV hull structure, vertical and horizontal plane data locations. Each individual point represents a CFD mesh vertex where the parameters have been recorded at. Tighter clustering is present in the boundary layer and wake region to adhere to the $y^+$ constraints for the turbulence model as detailed in Appendix \ref{App:HydrodynamicPhenomenon}.}
\label{fig:Domain_Plane_XLUUV}
\end{figure}

To illustrate the contents, quality and sampling density of the of the dataset, representative flow field visualisations are presented in a variety of different ways. Figure \ref{fig:Domain_Plane_XLUUV} details the vertical and horizontal planes as described. Also indicated in Figure \ref{fig:Domain_Plane_XLUUV} is the mesh density of the sampling points. This is higher in the wake region behind the XLUUV. Within this 3D domain, constrained by the two planes, the 3D data was also sampled in the regularly spaced grid. 

\begin{figure}[H]
    \centering
    \subfloat[]{\includegraphics[width=1.35in]{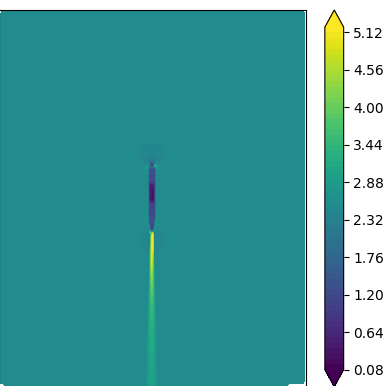}%
    \label{fig:VMag_00}}
    \hfil
    \subfloat[]{\includegraphics[width=1.35in]{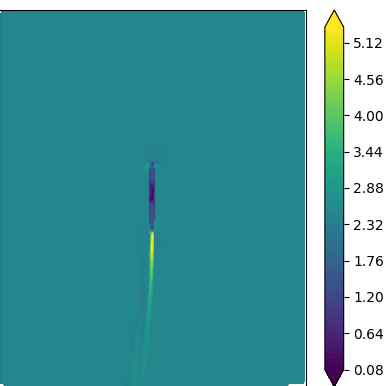}%
    \label{fig:VMag_10}}
    \hfil
    \subfloat[]{\includegraphics[width=1.35in]{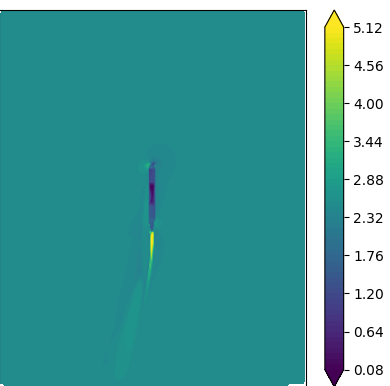}%
    \label{fig:VMag_20}}
    \hfil
    \subfloat[]{\includegraphics[width=1.35in]{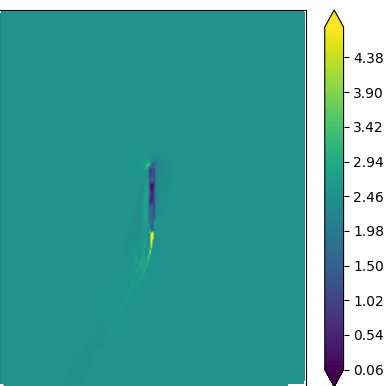}%
    \label{fig:VMag_30}}

    \caption{Horizontal planes with velocity magnitude contour plots shown. Velocity magnitude is provided in meters per second (m/s). Moving from left to right, a cross flow of \textbf{(a)} 0 degrees, \textbf{(b)} 10 degrees, \textbf{(c)} 20 degrees and \textbf{(d)} 30 degrees is shown. The XLUUV hull geometry is located in the centre of the planar slice. The wake structure is shown centre bottom of every planar slice.}
    \label{fig:VMag_Variation}
\end{figure}

\begin{figure}[H]
    \centering
    \subfloat[]{\includegraphics[width=1.35in]{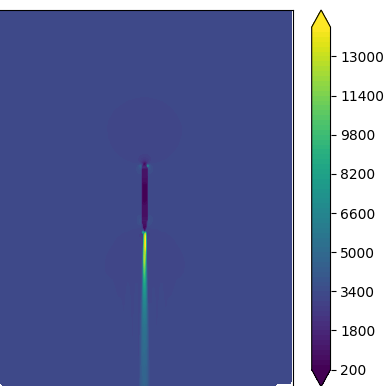}%
    \label{fig:DPress_00}}
    \hfil
    \subfloat[]{\includegraphics[width=1.35in]{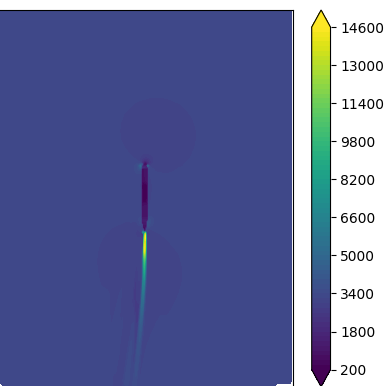}%
    \label{fig:DPress_10}}
    \hfil
    \subfloat[]{\includegraphics[width=1.35in]{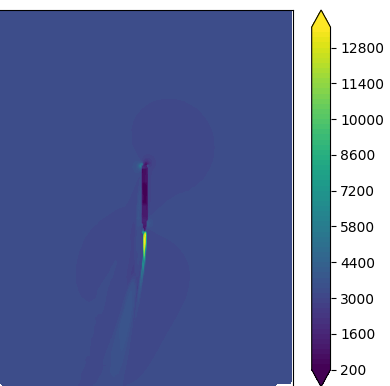}%
    \label{fig:DPress_20}}
    \hfil
    \subfloat[]{\includegraphics[width=1.35in]{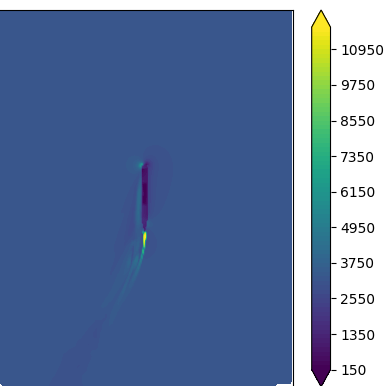}%
    \label{fig:DPress_30}}

    \caption{Horizontal planes with dynamic pressure contour plots shown. Dynamic pressure is provided in meters per second (Pa). Moving from left to right, a cross flow of \textbf{(a)} 0 degrees, \textbf{(b)} 10 degrees, \textbf{(c)} 20 degrees and \textbf{(d)} 30 degrees is shown. The XLUUV hull geometry is located in the centre of the planar slice. The wake structure is shown centre bottom of every planar slice.}
    \label{fig:DPress_Variation}
\end{figure}

\begin{figure}[H]
    \centering
    \subfloat[]{\includegraphics[width=1.35in]{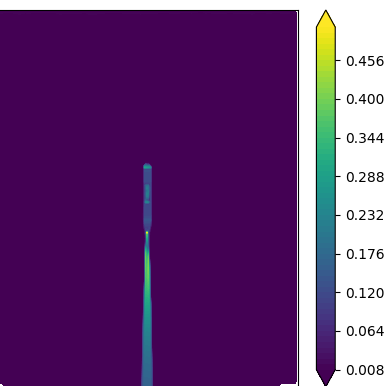}%
    \label{fig:TIntens_00}}
    \hfil
    \subfloat[]{\includegraphics[width=1.35in]{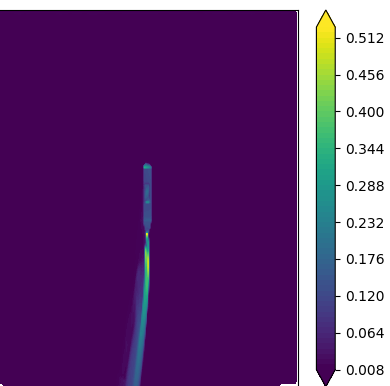}%
    \label{fig:TIntens_10}}
    \hfil
    \subfloat[]{\includegraphics[width=1.35in]{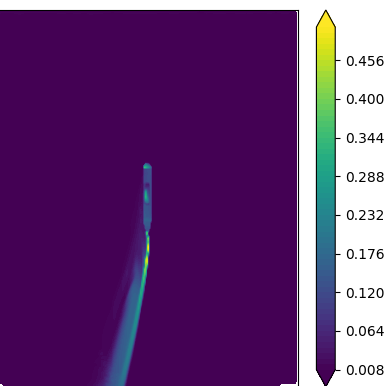}%
    \label{fig:TIntens_20}}
    \hfil
    \subfloat[]{\includegraphics[width=1.35in]{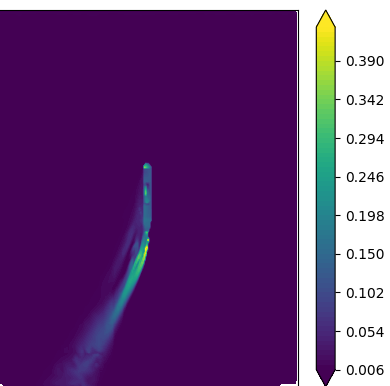}%
    \label{fig:TIntens_30}}

    \caption{Horizontal planes with turbulence intensity contour plots shown. Turbulence intensity is provided as a percentage (\%). Moving from left to right, a cross flow of \textbf{(a)} 0 degrees, \textbf{(b)} 10 degrees, \textbf{(c)} 20 degrees and \textbf{(d)} 30 degrees is shown. The XLUUV hull geometry is located in the centre of the planar slice. The wake structure is shown centre bottom of every planar slice.}
    \label{fig:TIntens_Variation}
\end{figure}

\begin{figure}[H]
    \centering
    \subfloat[]{\includegraphics[width=1.35in]{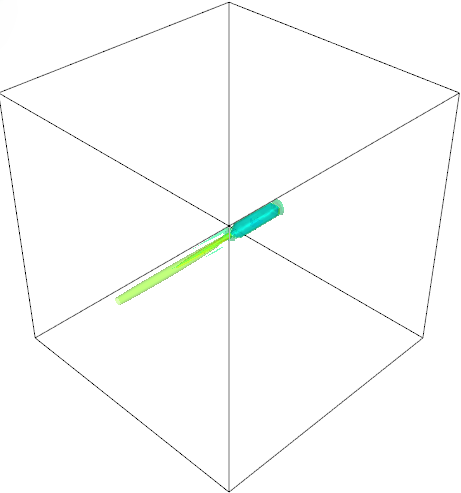}%
    \label{fig:VMag_3D_00}}
    \hfil
    \subfloat[]{\includegraphics[width=1.35in]{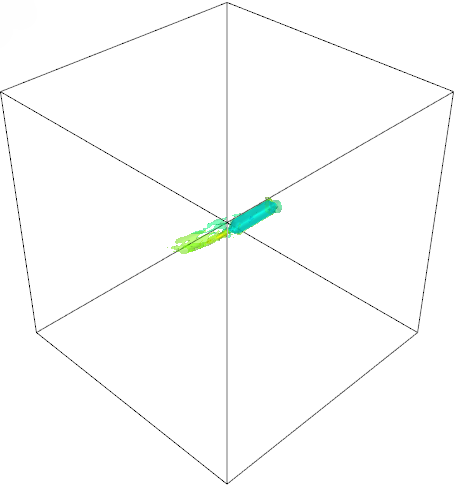}%
    \label{fig:VMag_3D_10}}
    \hfil
    \subfloat[]{\includegraphics[width=1.35in]{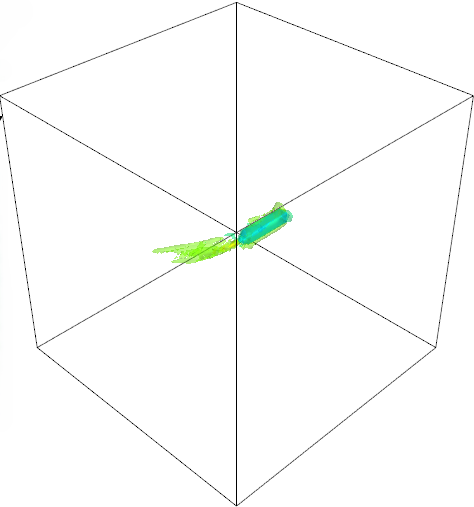}%
    \label{fig:VMag_3D_20}}
    \hfil
    \subfloat[]{\includegraphics[width=1.35in]{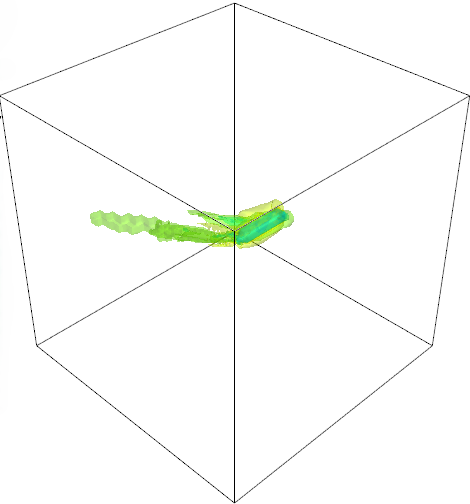}%
    \label{fig:VMag_3D_30}}

    \caption{3D volumes with velocity magnitude contour plots shown. Velocity magnitude is provided in meters per second (m/s). Moving from left to right, a cross flow of \textbf{(a)} 0 degrees, \textbf{(b)} 10 degrees, \textbf{(c)} 20 degrees and \textbf{(d)} 30 degrees is shown. The XLUUV hull geometry is located in the centre of the 3D domain. The wake structure is shown centre left of every volume.}
    \label{fig:VMag_3D_Variation}
\end{figure}

Furthermore, Figures \ref{fig:VMag_Variation}, \ref{fig:DPress_Variation} and \ref{fig:TIntens_Variation} are provided to give an indication of what the data looks like for the 2D slices. The horizontal plane is used to show the velocity (Figure \ref{fig:VMag_Variation}), dynamic pressure (Figure \ref{fig:DPress_Variation}) and turbulence intensity (Figure \ref{fig:TIntens_Variation}) in a cross-sectional view. This is shown around the XLUUV for different forward speeds and turning angles. An example of the 3D flow field structure is provided in Figure \ref{fig:VMag_3D_Variation}. This set of images details the 3D structure of velocity magnitude for the same cases as presented by Figure \ref{fig:VMag_Variation}.

These visualisations demonstrate the dataset's ability to capture a wide range of flow phenomena, including boundary layer development, wake formation, and the effects of turning manoeuvres. The detailed flow structures captured in the dataset are crucial for training ML models that can accurately predict the hydrodynamic forces and moments acting on underwater vehicles in complex flow conditions. The dataset goes far beyond the initial hydrodynamic analysis by encompassing a much broader range of speeds and turning angles, and by focusing on the generalised XLUUV model.

Preliminary exploration of the full dataset already reveals significant hydrodynamic trends. For example, it clearly visualises the evolution of the XLUUV's wake structure—including changes in wake width, length, and turbulence intensity—as a systematic function of both forward speed (and thus Reynolds number) and turning angle. The dataset also quantifies the degree of flow asymmetry and the formation of strong cross-flow induced vortical structures within and around the payload bay, particularly at higher turning angles and speeds. These comprehensive characteristics, capturing a wide spectrum of flow regimes from benign to highly complex, make the dataset exceptionally well-suited for training and validating ML models intended to predict such multi-parametric flow field evolutions, phenomena not typically captured in such detail or breadth in existing public datasets.

\subsection{Dataset Structure and Organisation}
\label{sec:DatasetStructure}

The WAKESET dataset is organised in a hierarchical structure designed to facilitate efficient access and processing for machine learning applications. This section provides detailed documentation of the data organisation, file formats, and usage guidelines.

\subsubsection{Directory Structure}
The dataset follows a standardised directory structure:

\begin{verbatim}
WAKESET/
|-- Volumes/
|   |-- Forward_0100_ms_Angle_00_CUBE_128/
|   |-- Forward_0100_ms_Angle_05_CUBE_128/
|   |-- ...
|-- Planes/
|   |-- Vertical/
|   |   |-- Forward_0100_ms_Angle_00_VERTPLN_ALL/
|   |   |-- Forward_0100_ms_Angle_05_VERTPLN_ALL/
|   |   |-- ...
|   |-- Horizontal/
|   |   |-- Forward_0100_ms_Angle_00_HORZPLN_ALL/
|   |   |-- Forward_0100_ms_Angle_05_HORZPLN_ALL/
|   |   |-- ...
|-- Examples/
|   |-- Python/
|   |   |-- requirements.txt
|   |   |-- README.md
|   |   |-- WAKESET_pytorch.py
|   |   |-- load_planes.py
|   |   |-- load_volumes.py
|   |   |-- load_visualizations.py
|-- README.md
|-- LICENSE
\end{verbatim}

\subsubsection{Coordinate System and Reference Frame}

The dataset uses a right-handed Cartesian coordinate system with the following conventions:
\begin{itemize}
    \item \textbf{Origin}: Located at the centre of the XLUUV's bow (forward-most point)
    \item \textbf{x-axis}: Points in the nominal forward direction of the XLUUV (positive aft)
    \item \textbf{y-axis}: Points to starboard (right side when facing forward)
    \item \textbf{z-axis}: Points upward (opposite to gravity)
    \item \textbf{Units}: All spatial coordinates are in meters
\end{itemize}

For turning maneuovers, the flow field is provided in the XLUUV body-fixed reference frame. The relationship between the body frame and the inertial frame is defined by the turning angle $\theta$ as described in Section \ref{sec:GeneralisationExpansion}.

\subsection{Data Augmentation}
\label{sec:DatasetAugmentation}

To further increase the size and diversity of the dataset, data augmentation techniques were employed. Data augmentation is a common and effective strategy in machine learning to artificially expand the dataset by applying transformations to the existing data \cite{AugmentedDataRotation, AugmentedDataGeneralisation}. In the context of CFD simulations, augmentation can help improve the generalisation ability of ML models and reduce the risk of over-fitting to specific flow conditions or orientations. It is particularly valuable when generating the original data is computationally expensive, as is the case with high-fidelity CFD.

For this dataset, the following augmentation techniques were applied:

\begin{enumerate}
    \item \textbf{Rotation:} The flow field data was rotated around the vertical axis (yaw) to generate new orientations while preserving the underlying flow physics. This is analogous to the rotation techniques used in image recognition \cite{AugmentedDataRotation}, where rotating an image does not change its fundamental content but provides a different perspective.  For each simulation with a non-zero turning angle, rotations were applied to create equivalent flow fields for the opposite turning direction.  For example, a simulation with a +10-degree turning angle was rotated to create a corresponding -10-degree turning angle case.

    \item \textbf{Flipping:} For simulations with a turning angle of 0 degrees (straight-ahead motion), the flow field data was mirrored across the vertical mid-plane (x-z plane). This takes advantage of the symmetry in the flow field for these cases, effectively doubling the number of samples for this subset of simulations. This is valid because the generalised XLUUV geometry is symmetric about the x-z plane.
\end{enumerate}

\begin{figure}[H]
    \centering
    \subfloat[]{\includegraphics[width=1in]{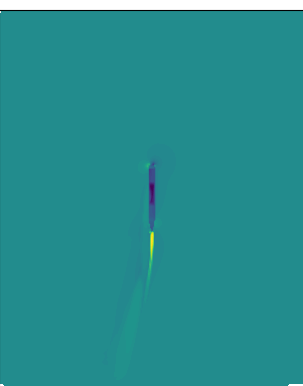}%
    \label{fig:VMag_-20_F}}
    \hfil
    \subfloat[]{\includegraphics[width=1in]{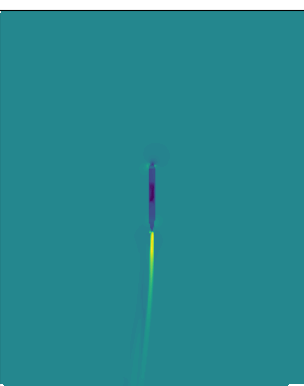}%
    \label{fig:VMag_-10_F}}
    \hfil
    \subfloat[]{\includegraphics[width=1in]{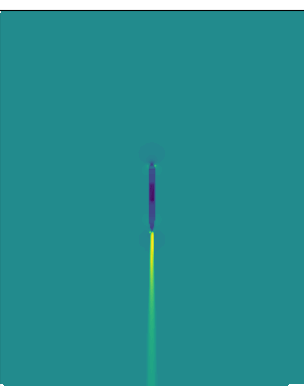}%
    \label{fig:VMag_00_F}}
    \hfil
    \subfloat[]{\includegraphics[width=1in]{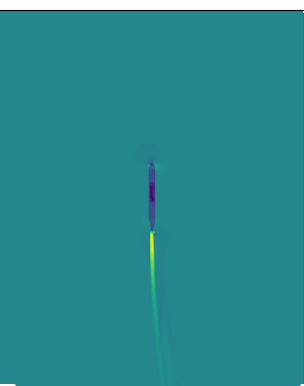}%
    \label{fig:VMag_10_F}}
    \hfil
    \subfloat[]{\includegraphics[width=1in]{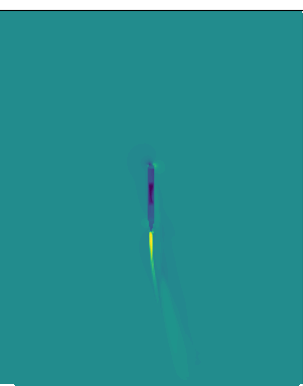}%
    \label{fig:VMag_20_F}}

    \caption{Horizontal planes of velocity magnitude shown with flipping augmentation applied. \textbf{(a)} 20 degrees, \textbf{(b)} 10 degrees, \textbf{(c)} 0 degrees, \textbf{(d)} -10 degrees and \textbf{(e)} -20 degrees are shown. \textbf{(d)} and \textbf{(e)} are flipped instances of \textbf{(a)} and \textbf{(b)}. The XLUUV hull geometry is located in the centre of the planar slice. The wake structure is shown centre bottom of every planar slice.}
    \label{fig:VMag_Variation_F}
\end{figure}

These transformations were applied to the velocity and pressure fields, as these are the primary quantities of interest for many ML applications in fluid mechanics.  The turbulent quantities (TKE, dissipation rate) were also transformed accordingly, ensuring consistency.  The transformations were implemented using custom scripts that manipulated the raw data files, ensuring that the augmented data remained physically consistent with the original simulations.

By applying the rotational and flipping transformations, the dataset size was augmented from 1,091 RANS flow simulations to 4,364, providing a four-fold increase in data. We provide a software utility to generate the augmented instances on-the-fly. The specific breakdown is as follows:

\begin{enumerate}
    \item \textbf{Original Dataset:} 1,091 simulations. 600 forward speed and angle combination pair simulations. 491 forward speed (dead ahead simulations). 

    \item \textbf{Augmented Dataset:} The 50 speeds x 12 turning angles are equal to 600 simulations, and a further 441 cases were generated with no turn angle. This totals 1041 instances. These are rotated to create the turning angle counterparts pairs via augmentation. Four sets of augmentations are done, with this yielding a total augmented dataset containing 4,364 instances.
\end{enumerate}

The augmented dataset offers several significant advantages for training ML models:

\begin{enumerate}
    \item \textbf{Improved Generalisation:} By exposing the models to a wider variety of flow orientations and conditions, the augmented dataset helps improve the models' ability to generalise to unseen scenarios. This is crucial for deploying ML models in real-world applications, where precise flow conditions may not be known a priori.

    \item \textbf{Enhanced Robustness:} The inclusion of rotated and flipped instances makes the models more robust to variations in the input data and reduces the likelihood of overfitting to specific orientations present in the original dataset \cite{AugmentedDataGeneralisation}.

    \item \textbf{Data Efficiency:} Augmentation allows for a more efficient use of the computational resources invested in generating the original dataset. By leveraging the existing simulations to create additional training instances, the need for running more computationally expensive CFD simulations is significantly reduced.
\end{enumerate}

The augmented dataset, with its increased size and diversity, serves as a valuable resource for training advanced ML models for fluid mechanics applications. It is expected that models trained on this dataset will exhibit improved prediction accuracy, generalisation ability, and robustness when applied to practical engineering problems involving complex underwater vehicle hydrodynamics.

\subsection{Data Availability}
\label{sec:DatasetAvailability}

To maximise the impact of this research and contribute to the advancement of ML applications in fluid mechanics, the generated dataset has been made publicly available to the research community \cite{HuggingFace}. The dataset, in the form detailed in Section \ref{sec:DatasetStructure} is hosted. This includes all data files (volumes and planes), loading functions for python use as well as PyTorch DataLoaders to integrated into ML workflows. The dataset is 480GB in size and can be accessed via a range of methods in the provided location \cite{HuggingFace}.

The documentation includes:
\begin{itemize}
    \item A detailed description of the dataset structure and organisation.
    \item Explanations of the variables included in the dataset (units, definitions).
    \item  Information on the CFD model used to generate the data (governing equations, turbulence model, boundary conditions, etc.).
    \item Details on the parametrisation and augmentation techniques employed.
    \item Instructions for accessing and using the dataset.
    \item Example scripts (e.g., in Python) for loading and processing the data.
    \item Contact information for inquiries and support.
    \item A suggested citation for the dataset.
\end{itemize}

By making the WAKESET dataset publicly available, this research aims to:

\begin{enumerate}
    \item \textbf{Facilitate Further Research:} The dataset can serve as a benchmark for developing and evaluating new ML architectures and training methodologies for fluid mechanics applications. Researchers can use the dataset to test their models and compare their performance against established methods.

    \item \textbf{Promote Reproducibility:} Researchers can use the dataset to reproduce the results presented in this paper and build upon the developed methodologies. This is a crucial aspect of scientific progress.

    \item \textbf{Encourage Collaboration:} The availability of the dataset can foster collaboration among researchers working on similar problems, leading to accelerated progress in the field. It can also facilitate interdisciplinary research, bringing together experts in fluid mechanics, machine learning, and data science.
\end{enumerate}

The dataset's availability aligns with the broader goal of addressing the scarcity of high-quality, large-scale CFD datasets for ML applications in fluid mechanics, as highlighted in Section \ref{sec:Motivation}. Beyond its sheer size and parameter range, initial observations from the dataset reveal, for instance, systematic transitions in wake morphology with varying Reynolds numbers and turning angles, and the significant distortion of the payload bay flow field under strong manoeuvring conditions. It is hoped that access to such detailed phenomena will serve as a valuable resource for the research community and contribute to the development of more accurate, efficient, and robust ML models for predicting complex hydrodynamic phenomena.

\section{Model Performance Benchmark}
\label{sec:Benchmarks}

The primary objective of WAKESET is to facilitate the development and benchmarking of machine learning models for fluid dynamics. To validate the dataset's utility and provide a reproducible baseline for future research, a conditional flow field generation benchmark was undertaken. This was conducted using a selection of prominent generative adversarial network (GAN) architectures. This benchmark demonstrates the viability of WAKESET for training data-intensive models and quantifies the initial performance for two key tasks: 2D flow slice prediction and 3D volumetric flow prediction. The models are tasked with predicting the flow field based on the input kinematic parameters (forward speed, $v_x$, and turning angle, $\theta$).

The benchmark was performed on each of the two distinct flow field prediction tasks. For the 2D flow slice generation, the model was conditioned on the input parameters ($v_x, \theta$) to predict a central $512 \times 512$ slice of the velocity magnitude field. This serves as a test for generating high-resolution flow features, such as the boundary layer and near-wake structures. For the 3D flow generation task, the model is conditioned on the input parameters to use the 2D slices ($512 \times 512$) and predict a $128 \times 128 \times 128$ volume of the velocity magnitude field. This task directly assesses the model's ability to capture complex, inherently three-dimensional flow structures and asymmetric flow patterns induced in the XLUUV's wake structure.

For the 2D task, three GAN architectures were selected to represent different subsets of the current GAN landscape. The first is the conditional deep convolutional GAN (cDCGAN) \cite{GAN_Flow1, GAN_Flow2}. The second is the Self-Attention GAN (SAGAN) \cite{zhang2019self} and the third is the PatchGAN \cite{isola2017image}. For the more computationally demanding task of 3D generation, the SAGAN, Wasserstein GAN with Gradient Penalty (WGAN-GP) \cite{WasserTraining, WasserGAN} and a custom 2D slice ingesting 2D3DGAN \cite{CFD_loss_3} were used.

\subsection{Experimental Setup and Data Preprocessing}

The entire augmented dataset of 4,364 instances was used for training and evaluation. To ensure the generalisability of the trained models across the entire parameter space, the dataset was split using a stratified sampling approach. The split was defined as 80\% for training, 10\% for validation and 10\% for testing, where the stratification was conducted with respect to speed and turning angle pairs across the three sets. This method was used to mitigate parameter-space overfitting and provides a robust measure of performance on unseen flow conditions.

The raw RANS simulation data was processed and interpolated onto structured grids to create standardised inputs for the ML models. For the 2D data, a central cross-sectional slice horizontally was taken. The mesh data was bi-linearly interpolated onto an equally spaced $512 \times 512$ grid. For the 3D data, the entire mesh domain was cubically interpolated onto a regularly spaced $128 \times 128 \times 128$ grid. In both cases, the target variable of velocity magnitude, was normalised using the min-max scaling into the range [0, 1]. This normalisation is a standard preprocessing step for generative networks and ensures stable training across the wide range of Reynolds numbers present in the dataset.

\subsection{Evaluation Metrics}

Model performance was assessed using a combination of image quality metrics and a physics-informed metric. The combination of the three provides a quantitative indication of the predicted flow fields with respect to both pixel-level accuracy and physical coherence. The performance was assessed via the Peak Signal-to-Noise Ratio (PSNR), Structural Similarity Index (SSIM), Fréchet Inception Distance (FID) and the relative error in the Area-Averaged Kinetic Energy ($\epsilon_{E_{k}}$.

The PSNR, as denoted by Eq. \ref{eqn:PSNR}, consists of the mean squared error (MSE) and the maximum possible pixel value ($MAX_I$) of the image ($I$). This relationship denotes the maximum possible power of corrupting noise that affects the fidelity of its perceived representation.

\begin{equation}
    \label{eqn:PSNR}
    PSNR = 10  \cdot  \log_{10} \left( \frac{MAX_I^2}{MSE} \right)
\end{equation}

The SSIM \cite{SSIM}, from Eq. \ref{eqn:SSIM}, is calculated between two windows of pixel values of the same size. In this formulation, $\mu_x$ denotes the pixel sample mean of $x$, $\mu_y$ the sample mean of $y$, $\sigma_x^2$ is the sample variance of $x$, $\sigma_y^2$ is the sample variance of $y$, $\sigma_{xy}$ is the sample covariance of $x$ and $y$. $c_1$ and $c_2$ are two variables introduced to stabilise the division with a weak denominator and are defined to be $c_1=(k_1 L)^2$, $c_2=(k_2 L)^2$ where $L$ is the dynamic range of the pixels and $k_1$ and $k_2$ are 0.01 and 0.03 respectively by default.

\begin{equation}
    \label{eqn:SSIM}
    SSIM(x, y) = \frac{(2 \mu_x \mu_y + c_1)(2 \sigma_{xy}+c_2)}{(\mu_x^2+\mu_y^2+c_1)(\sigma_x^2+\sigma_y^2+c_2)}
\end{equation}

The Fréchet Inception Distance (FID) \cite{FID_OG}, provided in Eq. \ref{eqn:FID}, is the standard metric used to assess the quality of images created by generative models, such as generative adversarial networks (GANs) and diffusion models. The FID compares the distribution of the generated images with the distribution of a set of real images (ground truths). A convolutional neural network, such as the inception architecture, produces high level features which are then compared. The Inception v3 model \cite{Szegedy_2016_CVPR} was used in this work. The FID compares two sets of images, as represented by two multidimensional Gaussian distributions $\mathcal{N}(\mu, \Sigma)$ and $\mathcal{N}(\mu', \Sigma')$. The distance between the two distributions is calculated as the 2-Wasserstein distance between the two distributions. Higher distances (FID scores) between the distributions indicate a poorer generative model, whilst conversely, lower scores indicate better performance. A score of 0 would indicate a perfect model.

\begin{equation}
    \label{eqn:FID}
    d_F(\mathcal{N}(\mu, \Sigma), \mathcal{N}(\mu', \Sigma'))^2 = ||\mu - \mu' || + tr(\Sigma + \Sigma' - 2(\Sigma \Sigma')^{1/2})
\end{equation}

To enforce physical consistency and evaluate the accuracy of the underlying flow physics the Relative Error in the Area-Averaged Kinetic Energy ($E_k$) is used. The area-averaged kinetic energy is calculated over the entire predicted 2D slice or 3D volume via Eq. \ref{eqn:Physics_1}. In this formulation $| \textbf{v}_i |$ is the predicted velocity magnitude at cell $i$, $\rho$ is the fluid density and $N$ is the number of cells. 

\begin{equation}
    \label{eqn:Physics_1}
    E_k = \frac{1}{N} \sum_{i=1}^{N} \frac{1}{2} \rho | \textbf{v}_i |^2
\end{equation}

The relative error, $\epsilon_{E_{k}}$, is then defined by Eq. \ref{eqn:Physics_2}. This metric quantifies the accuracy of the energy conservation quantity, and provides a robust, domain-specific measure of the 2D and 3D model's performance beyond the data driven visual similarity metrics. 

\begin{equation}
    \label{eqn:Physics_2}
    \epsilon_{E_{k}} = \frac{|E_{k,predicted} - E_{k,ground \ truth}|}{E_{k,ground \ truth}}
\end{equation}

The benchmark results for the two tasks are provided in Table \ref{tab:2D_Summary_Table}. All architectures were trained for a period of 200 epochs, with a batch size of 16 instances (2D or 3D). The training was undertaken in PyTorch on a dual NVIDIA RTX A6000 system, equipped with 256GB of RAM. In training, an epoch-batch loop was implemented. The Adam optimiser was used, and the GAN Binary Cross Entropy (BCE) loss was augmented with a structural flow metric. This was the Gradient Mean Squared Error (GMSE) loss \cite{cooper2024generalised}. This loss is specifically designed to operate on computational fluid dynamics data, such as the WAKESET. In addition to the metrics above, are the inference time $t$ and the model size upon inference $M_s$ to give an indication of the computational efficiency. 

















\begin{table}[H]
\caption{Performance comparison of generative models on the WAKESET benchmark. Metrics include Peak Signal-to-Noise Ratio (PSNR), Structural Similarity (SSIM), Fréchet Inception Distance (FID), relative error in area-averaged kinetic energy ($\epsilon_{E_{k}}$), inference time ($t$), and model size ($M_s$). Arrows indicate direction of better performance.}
\label{tab:2D_Summary_Table}
\begin{center}
\footnotesize
\begin{tabular}{ll cccccc}
\toprule \toprule
\textbf{Task} & \textbf{Model} & \textbf{PSNR} $\uparrow$ & \textbf{SSIM} $\uparrow$ & \textbf{FID} $\downarrow$ & $\epsilon_{E_{k}}$ $\downarrow$ & $t$ ($\mu s$) & $M_s$ (GB) \\
\midrule
\multirow{3}{*}{2D [$512^2$]} 
& cDCGAN \cite{GAN_Flow1} & \textbf{47.24} & \textbf{0.998} & \textbf{43.01} & \textbf{0.015} & 4.51 & \textbf{0.422} \\
& SAGAN \cite{zhang2019self} & 18.70 & 0.536 & 329.51 & 0.283 & 15.80 & 1.203 \\
& PatchGAN \cite{isola2017image} & 46.26 & 0.997 & 54.43 & 0.018 & 4.51 & 0.423 \\
\midrule
\multirow{3}{*}{3D [$128^3$]} 
& SAGAN \cite{zhang2019self} & 29.06 & 0.602 & \textbf{27.99} & \textbf{0.081} & 95.97 & 10.39 \\
& WGAN-GP \cite{WasserGAN} & 15.62 & 0.478 & 341.98 & 0.312 & 14.08 & 1.21 \\
& 2D3DGAN \cite{CFD_loss_3} & \textbf{31.62} & \textbf{0.752} & 285.53 & 0.113 & \textbf{1.48} & \textbf{0.306} \\
\bottomrule \bottomrule
\end{tabular}
\end{center}
\end{table}

The 2D benchmark demonstrates a high reconstruction quality for two of the three models. The cDCGAN achieved the highest $\epsilon_{E_{k}}$ accuracy (1.5\% relative error), while PatchGAN exhibited comparable PSNR and SSIM, indicating excellent pixel-level correspondence. This suggests that for predicting 2D slices where the overall structure is highly constrained by the boundary conditions, simpler convolutional models are capable of achieving high fidelity, likely benefiting from the straified sampling and the dataset's clear parameter-to-flow mapping. Conversely, the SAGAN model struggled significantly with the $512^2$ resolution, achieving the lowest PNSR and highest FID. Its high $M_s$ and $t$ further confirm the overhead of the self-attention mechanism did not provide a corresponding benefit in structural capture for this specific 2D task.

The 3D prediction task is inherently more challenging due to the significantly increased dimensionality and the complexity of predicting volumetric flow features. The performance metrics reflect this difficulty, with generally lower PSNR and SSIM scores compared to the 2D task as shown in Table \ref{tab:2D_Summary_Table}.

The 3D SAGAN architecture however, emerged as the most efficient and physically accurate model, achieving the best $\epsilon_{E_{k}}$ (8.1\% relative error) and the lowest FID, proving the benefit of the self-attention mechanism in such 3D tasks. Whereby it is better suited to capture the volumetric distribution and perceptual realism of the flow. The WGAN-GP model provided the poorest performance across all metrics, highlighting its limitation in generating highly structured, volumetric data from a large parameter space. The 2D3DGAN emerged as an efficient architecture, given its computational footprint and highest scoring PSNR and SSIM, as well as the lowest inference time.

\section{Discussion}
\label{sec:Discussion}

The benchmark results confirm that the WAKESET dataset is a suitable and challenging resource for training deep learning models for fluid mechanics in a data-driven capacity. The results here establish a performance baseline, demonstrating that while simple architectures can achieve high reconstruction accuracy in 2D, the complexity of 3D flow fields and high Reynolds numbers can differentiate more specialised architectures (such as the 2D3DGAN and the structure-focused SAGAN) allowing for models to be compared in a comprehensive and informative manner. 

The development of a novel, large-scale, high-fidelity CFD dataset is specifically designed to address the critical need for comprehensive training data in machine learning applications for fluid mechanics that are highly turbulent, sufficiently large and focused on a practical engineering problem. The dataset was developed based on the complex hydrodynamics of underwater vehicle interactions, using the scenario of an AUV recovering within an XLUUV. The methodology began with a detailed hydrodynamic analysis of this specific scenario \cite{ZacIEEE}, then before it was systematically generalised and expanded from this initial investigation to create a versatile resource for training a variety of ML models (Section \ref{sec:Generalisation}) by ensuring relevant physical effects and flow structures were captured. The dataset development process demonstrated:

It is observed that the new dataset offers the second highest number of training instances \cite{Huang}, the highest upper Reynolds number \cite{Ahmed_2016}, and comprehensive 3D data, features that significantly enhance its applicability for training advanced ML architectures in fluid mechanics. The large number of instances (4,364 after augmentation) allows for the training of more complex, data-intensive ML models (e.g., deep neural networks with high parameter counts) whilst mitigating the risk of overfitting and potentially promoting better generalisation to unseen flow conditions. The high Reynolds limit (up to $1.09 \times 10^8$) ensures that the dataset represents highly turbulent flows, characteristic of many full-scale engineering applications. This enables ML models to learn features relevant to these challenging, practical regimes, which are often underrepresented in existing datasets focused on lower Re or canonical flows. The 3D nature of the data is crucial for capturing the inherently three-dimensional physics of most real-world fluid dynamic problems. Including complex vortical structures, flow separation, and anisotropic turbulence. This allows ML models to learn and predict these complex spatial features, which are often lost or oversimplified in 2D datasets \cite{MegaFlow2D, McConkey, AirfRANS, Huang}. Collectively, these attributes position the dataset as a valuable resource for developing and benchmarking ML models capable of handling the complexities of practical engineering flows.

\section{Conclusion}
\label{sec:Conclusion}

This paper details the creation of a unique and valuable resource for the fluid mechanics ML community. The dataset's combination of high-fidelity CFD simulations, generalised geometry, extensive parametrisation, and data augmentation provides a robust foundation for training and evaluating ML models capable of accurately predicting complex underwater vehicle hydrodynamics.  This addresses a significant gap in existing resources, which are often limited in size, scope, and diversity (as discussed in Section \ref{sec:Motivation}).  Future research directions may include expanding the dataset to include transient simulations, incorporating additional geometric variations, and exploring the application of the dataset to a wider range of ML architectures and tasks.  Overall, this dataset stands as a significant contribution, enabling the development of more accurate, efficient, and generalisable ML models for fluid mechanics applications, and bridging the gap between computational fluid dynamics and data-driven modelling.

\section*{Acknowledgements}

This dataset was developed with the assistance and resources from the National Computational Infrastructure (NCI) which is supported by the Australian Government, conducted as part of the Australian HPC-AI Talent Program. 

\bibliographystyle{unsrturl}
\bibliography{refs}

\newpage

\appendix

\section{Hydrodynamic Phenomenon of Interest}
\label{App:HydrodynamicPhenomenon}

Several key hydrodynamic phenomena were analysed to understand their impact on the AUV during the recovery process. These phenomena are crucial for determining the physical effects the flow field exerts on the AUV. These are often targeted values for ML flow field applications where the drag and lifting forces on a body of interest are of high importance - due to them defining how the body is effected by the fluid medium. In this initial investigation, the following quantities were investigated:

The lift and drag forces acting on the AUV are critical for assessing its ability to maintain the desired trajectory and achieve robust and safe docking with the XLUUV. The lift and drag forces describe how the AUV will be pushed in the vertical and lateral directions and define the resistive forces it must overcome. These forces were non-dimensionalised into lift ($C_{L}$) and drag ($C_{D}$) coefficients to provide a basis for comparison across different velocities. The lift coefficient, $C_{L}$, was determined with respect to the vertical direction. $C_{L}$ is a function of the lift force ($F_{L}$), fluid density ($\rho$), velocity ($v$), and the projected area in the y-axis direction ($A_{L}$), as given by Eq. \ref{eqn:LiftCoeff} \cite{42}. The drag coefficient, $C_{D}$, is determined with respect to the horizontal direction. $C_{D}$ is a function of the drag force ($F_{D}$), fluid density ($\rho$), velocity ($v$), and the projected area in the x-axis direction ($A_{D}$), as given by Eq. \ref{eqn:DragCoeff} \cite{42}.

\begin{equation}
    \label{eqn:LiftCoeff}
    C_L =\frac{2 F_L}{\rho v^2 A_L}
    \end{equation}

    \begin{equation}
    \label{eqn:DragCoeff}
    C_D =\frac{2 F_D}{\rho v^2 A_D}
    \end{equation}

Turbulent flow effects can significantly impact the performance, control, and navigation of smaller underwater vehicles \cite{44}. To quantify these effects, three components were assessed. The first is the turbulence intensity ($I$). This indicates the magnitude of velocity fluctuations relative to the freestream velocity. It is defined as the ratio of the root-mean-square of the velocity fluctuations ($v'$) to the freestream velocity ($v_{free}$), as given by \ref{eqn:TurbIntens} \cite{45}.

            \begin{equation}
            \label{eqn:TurbIntens}
            I = \frac{v'}{v_{free}}
            \end{equation}

Second is the turbulent kinetic energy (TKE). This represents the mean kinetic energy per unit mass associated with eddies in turbulent flow, providing an indication of the energy contained within the turbulent flow. It is defined as half the sum of the variances of the velocity components ($v_{x}, v_{y}, v_{z}$), as given by Eq. \ref{eqn:TurbKinEn} \cite{46}.

            \begin{equation}
            \label{eqn:TurbKinEn}
            k \equiv \frac{1}{2} \left [ \overline{v'^{2}{x}} + \overline{v'^{2}{y}} + \overline{v'^{2}_{z}} \right ]
            \end{equation}

Third is the vorticity magnitude. This indicates the speed at which the flow is rotating, with higher magnitudes indicating faster swirling structures. Vorticity ($\vec{w}$) is defined as the curl ($\bigtriangledown$) of the flow velocity vector ($\vec{\textbf{v}}$), as given by Eq. \ref{eqn:VortMag} \cite{47}.

            \begin{equation}
            \label{eqn:VortMag}
            \vec{w} = \bigtriangledown \times \vec{\textbf{v}}
            \end{equation}

To provide a comprehensive understanding of the turbulent flow effects impacting the AUV, these quantities were calculated as area-weighted averages over the AUV's surface, as given by Eq. \ref{eqn:AreaWAvg} \cite{46}, where $A$ is the total area of the AUV surface and $\phi$ is the variable of interest.

\begin{equation}
\label{eqn:AreaWAvg}
\frac{1}{A} \int \phi dA = \frac{1}{A} \sum_{i=1}^{n} \phi_{i} \left | A_{i} \right |
\end{equation}

These hydrodynamic quantities were carefully selected to provide a detailed characterisation of the flow field and its impact on the AUV during the recovery process. By analysing these quantities across different approach paths, trajectories, and speeds, the initial analysis captures a wide range of hydrodynamic phenomena, providing a solid foundation for the subsequent dataset generalisation.

\section{Hydrodynamic Model Generalisation}
\label{App:HydrodynamicModelGeneralisation}

The CFD model used for generating the large-scale dataset builds upon the model described in \cite{ZacIEEE}, with adaptations made to accommodate the wider range of parameters (Section \ref{sec:GeneralisationExpansion}) and to ensure computational efficiency for the large number of simulations required. The computational domain was designed using the characteristic length (L) of the generalised XLUUV, which is 22 meters. The domain extends 3L (66 meters) ahead of the XLUUV's bow, 5L (110 meters) to the sides, top, and bottom, and 8L (176 meters) behind the XLUUV's stern. These dimensions ensure that the boundaries do not unduly influence the flow field around the XLUUV and that the wake is adequately captured, even for the expanded range of flow conditions.

Within the domain, several bodies of influence (BOIs) were defined for localised mesh refinement. These regions were identified in the prior hydrodynamic analysis where flow conditions are complex and need to be fully resolved in an accurate manner. This is achieved through the use of tighter discretization in this region via the BOIs and associated mesh controls. These regions are primarily where the wake and payload bay flow structures exist. A payload bay BOI extends 3 meters from the front of the bay and 6 meters to the sides and rear, ensuring that the complex flow features in this critical region are accurately resolved. A circular BOI with a diameter of 5 meters extends 3 meters ahead of the propeller and 10 meters astern, capturing the flow into and out of the propeller. Two smaller, secondary BOIs are located within the primary domain to further refine the mesh around the XLUUV body geometry and propeller wake structures more broadly. These BOIs are consistent with the initial analysis, but their effectiveness is re-evaluated for the generalised model and expanded parameter space via post processing of the solution fields after calculation.

To accurately model the influence of the XLUUV's propeller on the flow field, an actuator disk model was incorporated, with its parameters varied inline with the parameter spaces velocity and turn angle metrics. This actuator disk model simulates the effects of the propeller without the need to resolve the complex blade geometry, significantly reducing computational cost on a per simulation basis. The propeller model was developed based on the relationship between two commercial propellers, the T-200 and T-500 \cite{2_22}, scaled up to match the size and thrust requirements of the generalised XLUUV. This represents a departure from the initial analysis, which used the INSEAN E1619 propeller model. The change was made to ensure the propeller model was representative of commercially available options and to facilitate scaling to the generalised XLUUV.

The required thrust was estimated by first calculating the drag force acting on the XLUUV at the maximum speed of 5 m/s. Using the drag coefficient ($C_{D,prev}$) derived from a previous study \cite{ZacIEEE} involving a similar XLUUV design, and applying it to the generalised XLUUV model, the drag force was estimated as:

\begin{equation}
\label{eqn:DragCoeffApplied}
C_{D, prev} = \frac{F_{Drag}}{\frac{1}{2} \cdot \rho \cdot v^{2} \cdot S} = \frac{(5880)}{\frac{1}{2} \cdot (1025.1627) \cdot (1.55)^{2} \cdot (6.76)} = 0.706
\end{equation}

\begin{equation}
\label{eqn:DragForceApplied}
F_{D, new} = \frac{1}{2} \cdot \rho \cdot v^{2} \cdot C_{D} \cdot S = \frac{1}{2} \cdot (1025.1627) \cdot (5.00)^{2} \cdot (0.706) \cdot (4) = 36,223.54 N
\end{equation}

A 5\% safety factor was applied to the pressure jump calculation to account for propulsion losses and ensure sufficient thrust. The pressure jump was then calculated by dividing the required thrust by the blade area (1.0m outer, 0.35m hub, resulting in an area of 0.68918 $m^{2}$), yielding a value of 52,559.83 Pa, which was then adjusted to 55,187 kPa after applying the safety factor.

The propeller tip speed ($V_{tip}$) and pressure jump ($P_{J}$) were varied based on the incoming flow speed to the domain to ensure that the thrust matched the drag for each simulation, accurately modelling the underway state of the vessel. The relationship between tip speed and domain speed ($v_{d}$) is given by:

\begin{equation}
\label{eqn:PropellerTipSpeed}
V_{tip} = \frac{\pi \cdot D \cdot n_{rpm}}{60}
\end{equation}

\begin{equation}
\label{eqn:PropellerTipSpeedEQN}
V_{tip} = 4.620 \cdot v_{d}
\end{equation}

The relationship between pressure jump ($P_{J}$) and domain speed ($v_{d}$) is given by:

\begin{equation}
\label{eqn:PressureJumpEQN}
P_{J} = [1997.6 \cdot v_{d}^{2}] - [6 \cdot 10^{-10} \cdot v_{d}^{2}] + [7 \cdot 10^{-10}]
\end{equation}

These relationships were implemented in the CFD model to ensure that the propeller's performance was accurately represented across the full range of simulated flow conditions, including the expanded speed range. Table \ref{Tab:PropellerDataTable} summarises the key parameters of the developed propeller model.

\begin{table}[!ht]
\caption{Propeller characteristics determined from the INSEAN E1619 propeller design and the work provided by \cite{ZacIEEE, SUBSTEC}, adapted to the generalised XLUUV design for the dataset.}
\label{Tab:PropellerDataTable}
\begin{center}
\footnotesize
\begin{tabular}{lll}
\toprule \toprule
\textbf{Parameter} & \textbf{E1619} & \textbf{Dataset XLUUV} \\ \toprule \toprule
Propeller diameter & 0.485 [$m$] & 1.25 [$m$] \\ \hline
Inflow velocity & 1.68 [$m/s$] & 5.00 [$m/s$] \\ \hline
Blade tip velocity & 6.21 [$m/s$] & 18.48 [$m/s$] \\ \hline
Rotational frequency & 4.08 [$Hz]$ & 5.88 [$Hz$] \\ \hline
Advance ratio & 0.85 & 0.85 \\ \hline
Required thrust & - & 36,223 [$N$] \\ \hline
Blade (max dia.) & 0.485 [$m$] & 1.00 [$m$] \\ \hline
Blade (hub dia.) & - & 0.35 [$m$] \\ \hline
Blade area & - & 0.68918 [$m^{2}$] \\ \hline
N/Blade area & - & 52,559.83 [$N/m^{2}$] \\ \hline
Safety factor & - & 1.05 \\ \hline
Pressure jump & - & 55.188 [$kPa$]\\ \toprule \toprule
\end{tabular}
\end{center}
\end{table}

While the initial hydrodynamic analysis \cite{ZacIEEE} established the fundamental CFD methodology and identified key flow phenomena and sensitive regions (payload bay, propeller wake, boundary interaction zones), the transition to a generalised XLUUV geometry (Section \ref{sec:GeneralisationGeometry}) and a significantly expanded operational parameter space (Section \ref{sec:GeneralisationExpansion}) necessitated a new, dedicated mesh convergence study for the dataset generation phase. This re-evaluation was crucial because alterations in hull form, payload bay dimensions and the broader range of Reynolds numbers and flow angles could introduce different local flow gradients and thus alter the mesh resolution requirements for converged and accurate solutions. The findings from the initial analysis, particularly regarding the locations of complex flow structures, directly informed the strategy for localised mesh refinement (BOIs) in this new study. However, the specific cell sizing and overall mesh count required independent verification for the generalised model to ensure the accuracy of the final dataset. This study involved creating a series of meshes with varying cell counts from $1.382 \times 10^6$ to $38.326 \times 10^6$ cells, with details provided in Table \ref{Tab:MeshDatasetSweep}.

\begin{table}[!ht]
\caption{Mesh parameters used to conduct the convergence analysis for the CFD simulation environment for the dataset development. The table indicates the mesh area and body of influence (BOI) sizings, in addition to domain size controls and mesh metrics.}
\label{Tab:MeshDatasetSweep}
\begin{center}
    \footnotesize
        \begin{tabular}{lllllllll}
        \toprule \toprule
        \textbf{Area}  &  \textbf{Metric} & \textbf{1}  & \textbf{2} & \textbf{3} & \textbf{4} & \textbf{5} & \textbf{6} & \textbf{7}

        \\ \toprule \toprule
        BOI & Prop.     & 0.04 & 0.03 & 0.02 & 0.015 & 0.015 & 0.015 & 0.015 \\
         & Wake.        & 0.45 & 0.35 & 0.30 & 0.25 & 0.25 & 0.25 & 0.25 \\
         & Wake.        & 0.25 & 0.15 & 0.10 & 0.08 & 0.08 & 0.06 & 0.05 \\
         & Wake.        & 0.15 & 0.10 & 0.08 & 0.07 & 0.05 & 0.04 & 0.03 \\
         & Payload bay  & 0.15 & 0.10 & 0.08 & 0.07 & 0.05 & 0.04 & 0.04 \\ \hline

        Domain & Min.   & 0.07 & 0.06 & 0.05 & 0.04 & 0.03 & 0.02 & 0.018 \\
         & Max.         & 3.00 & 3.00 & 3.00 & 3.00 & 3.00 & 3.00 & 3.00 \\
         & Growth rate  & 1.20 & 1.20 & 1.20 & 1.20 & 1.20 & 1.20 & 1.20 \\
         & Curv. Ang.   & 18   & 18   & 18   & 18   & 18   & 18   & 18  \\
         & C.P.G.       & 1    & 1    & 1    & 1    & 1    & 1    & 1   \\ \hline

        Mesh & $\#$ Layers. & 12  & 14 & 16 & 18 & 20 & 22 & 24 \\
         & Growth rate      & 1.20 & 1.20 & 1.20 & 1.20 & 1.20 & 1.20 & 1.20 \\
         & $y^{1}$ [$\times 10^{-3}$] & 1.59 & 1.59 & 1.59 & 1.59 & 1.59 & 1.59 & 1.59 \\
         & Peel layers      & 2 & 2 & 3 & 3 & 4 & 4 & 4 \\
         & Min cell length  & 0.04 & 0.03 & 0.03 & 0.02 & 0.02 & 0.015 & 0.013\\
         & Max cell length  & 1.60 & 1.60 & 1.60 & 1.60 & 1.60 & 1.60 & 1.60\\ \hline

         Quality & Min quality  & 0.078 & 0.108 & 0.100 & 0.094 & 0.068 & 0.054 & 0.051 \\
         & Total cells [$\times 10^{6}$]   & 1.382 & 3.080  & 6.507  & 9.416 & 14.475  & 25.976  & 38.326 \\ \toprule \toprule
        \end{tabular}
        \end{center}
    \end{table}

\begin{figure}[!ht]
    \centering
    \subfloat[]{\includegraphics[width=2.2in]{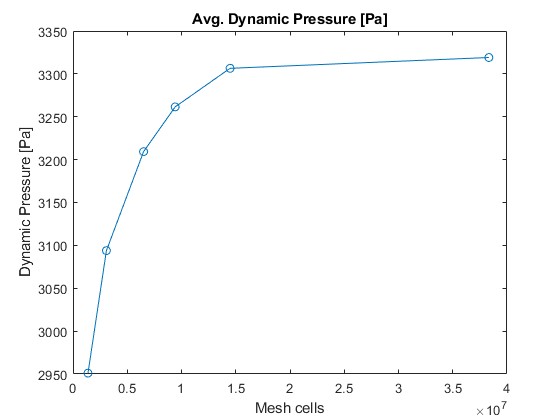}%
    \label{fig:Mesh_DynamicPressure}}
    \hfil
    \subfloat[]{\includegraphics[width=2.2in]{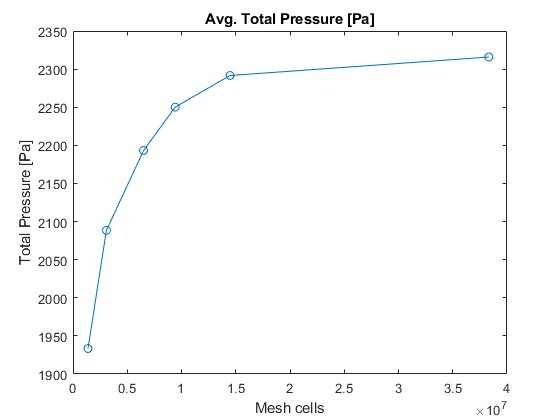}%
    \label{fig:Mesh_TotalPressure}}
    \\
    \subfloat[]{\includegraphics[width=2.2in]{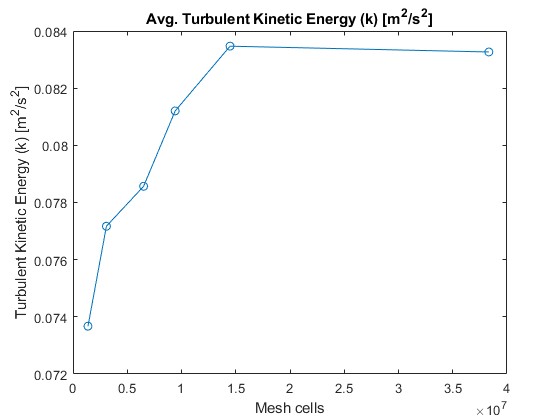}%
    \label{fig:Mesh_TKE}}
    \hfil
    \subfloat[]{\includegraphics[width=2.2in]{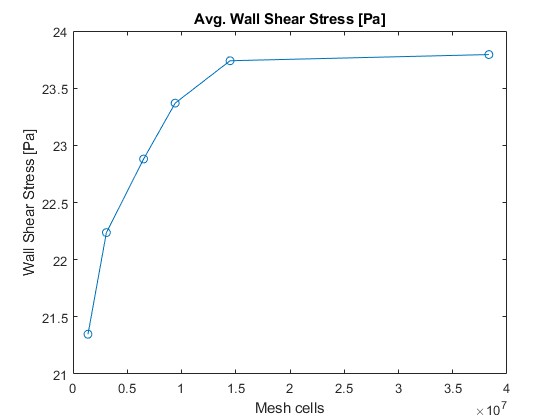}%
    \label{fig:Mesh_WallShear}}
    \\
    \subfloat[]{\includegraphics[width=2.2in]{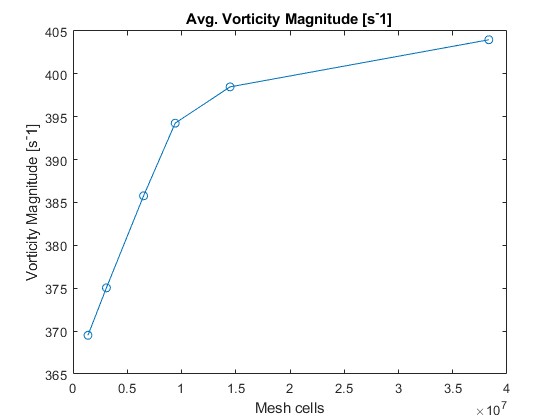}%
    \label{fig:Mesh_Vorticity}}
    \caption{Mesh convergence data for the facet average (a) dynamic pressure, (b) total pressure, (c) turbulent kinetic energy, (d) wall shear, (e) vorticity.}
    \label{fig:DatasetMeshConvergence}
\end{figure}

The convergence of key parameters, including dynamic pressure, total pressure, turbulent kinetic energy, wall shear, and vorticity, was assessed across the different mesh sizes. The results, presented in \ref{fig:DatasetMeshConvergence}, demonstrate that convergence is achieved as the cell count approaches 10-15 million cells.

Based on the new convergence analysis, a mesh size of 14 million cells was selected for the dataset generation. This mesh size provides a balance between computational cost and solution accuracy, for the RANS simulations, as indicated by the convergence of the key parameters (\ref{fig:DatasetMeshConvergence}). The selected mesh ensures that the dataset accurately captures the essential flow features while remaining computationally feasible for the large number of simulations required. This choice is justified independently of the initial analysis, demonstrating rigour.

\begin{figure}[!ht]
    \centering
    \subfloat[]{\includegraphics[width=2.2in]{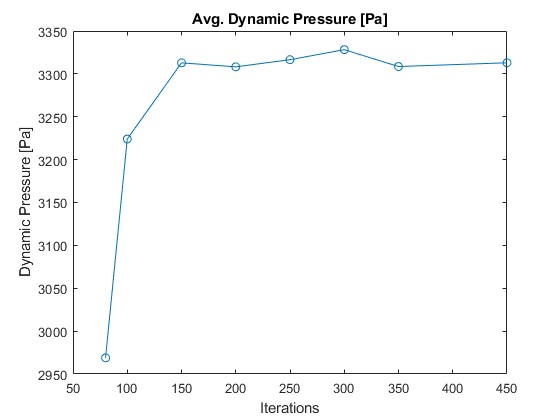}%
    \label{fig:Iter_DynamicPressure}}
    \hfil
    \subfloat[]{\includegraphics[width=2.2in]{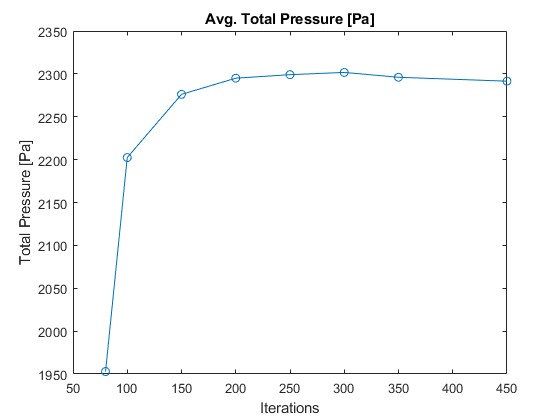}%
    \label{fig:Iter_TotalPressure}}
    \\
    \subfloat[]{\includegraphics[width=2.2in]{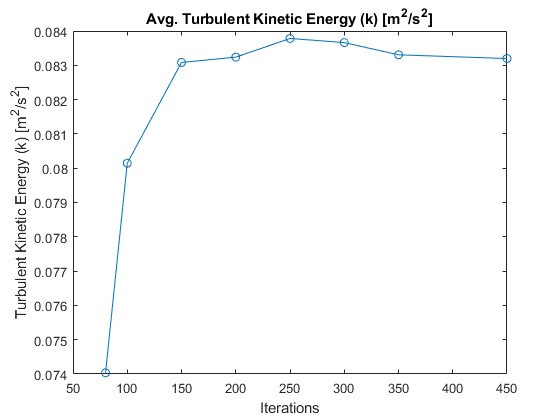}%
    \label{fig:Iter_TKE}}
    \hfil
    \subfloat[]{\includegraphics[width=2.2in]{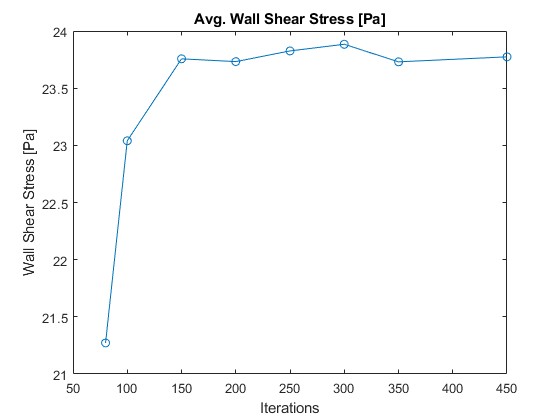}%
    \label{fig:Iter_WallShear}}
    \\
    \subfloat[]{\includegraphics[width=2.2in]{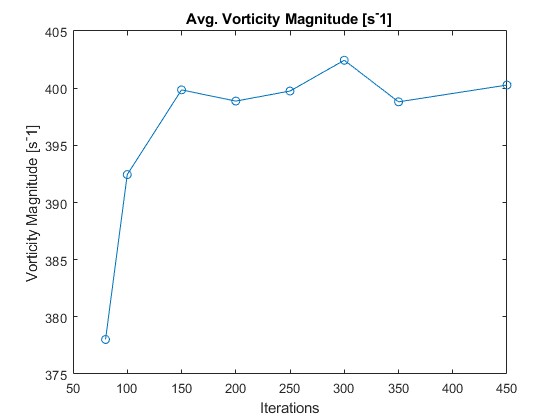}%
    \label{fig:Iter_Vorticity}}
    \caption{Iteration convergence data for the facet average (a) dynamic pressure, (b) total pressure, (c) turbulent kinetic energy, (d) wall shear, (e) vorticity.}
    \label{fig:DatasetIterConvergence}
\end{figure}

In addition to mesh convergence, an iteration convergence study was performed to determine the appropriate number of iterations required for each simulation to reach a converged state. This study was also conducted specifically for the generalised model and expanded parameter space, ensuring its validity for the dataset generation. The study focused on the same key parameters as the mesh convergence analysis: dynamic pressure, total pressure, turbulent kinetic energy, wall shear, and vorticity. The results, presented in \ref{fig:DatasetIterConvergence}, demonstrate that convergence is achieved for all parameters between 200 and 250 iterations.

Based on the iteration convergence study, a conservative cutoff of 300 iterations was selected for each simulation in the dataset. This ensures that all simulations reach a well-converged state, minimising numerical errors and ensuring the reliability of the generated data. The generation of the dataset required substantial computational resources due to the large number of simulations (1,091), the complex flow physics involved, and the need for fine mesh resolution and a high number of iterations. To meet these demands, the simulations were performed using the GADI supercomputer at the National Computational Infrastructure (NCI) in Australia \cite{NCIRef}. The use of GADI's high-performance computing capabilities was crucial for completing the dataset generation within a reasonable time-frame. This highlights the importance of access to advanced computational resources for generating large, high-fidelity CFD datasets.

\section{WAKESET Distribution Metrics}
\label{App:WAKESET_Distribution_Metrics}

\end{document}